\documentclass[journal]{IEEEtran}
\usepackage{cite}
\usepackage{amsmath,amssymb}
\usepackage{algorithm,algpseudocode}
\usepackage{graphicx,color, upgreek}
\usepackage{bbm}
\usepackage{subfigure}

\makeatletter
\def\endthebibliography{%
  \def\@noitemerr{\@latex@warning{Empty `thebibliography' environment}}%
  \endlist
}
\makeatother

\begin{document}
\bstctlcite{IEEE:BSTcontrol}

\title{Extended Version of ``New Theory and Faster Computations for Subspace-Based Sensitivity Map Estimation in Multichannel MRI'' }
\author{Rodrigo A. Lobos, \IEEEmembership{Member, IEEE}, Chin-Cheng Chan, Justin P. Haldar, \IEEEmembership{Senior Member, IEEE}
\thanks{This work was supported in part by NIH research grants R01-MH116173 and  R01-NS074980.}
\thanks{R. Lobos, C.-C. Chan, and J. Haldar are with the Signal and Image Processing Institute, Ming Hsieh Department of Electrical and Computer Engineering, University of Southern California, Los Angeles, CA, 90089 USA. }}

\maketitle 
 
\begin{abstract}
This is an unabridged version of a journal manuscript that has been submitted for publication \cite{lobos2023piscoJ}.  (Due to length restrictions, we were forced to remove substantial amounts of content from the version that was submitted to the journal, including more detailed theoretical explanations, additional figures, and a more comprehensive bibliography.  This content remains intact in this version of the document).

Sensitivity map estimation is important in many multichannel MRI applications.  Subspace-based sensitivity map estimation methods like ESPIRiT are popular and perform well, though can be computationally expensive and their theoretical principles can be nontrivial to understand.  In the first part of this work, we present a novel theoretical derivation of subspace-based sensitivity map estimation based on a linear-predictability/structured low-rank modeling perspective.  This results in an estimation approach that is equivalent to ESPIRiT, but with distinct theory that may be more intuitive for some readers.  In the second part of this work, we propose and evaluate a set of computational acceleration approaches (collectively known as PISCO) that can enable substantial improvements in computation time (up to $\sim$100$\times$ in the examples we show) and memory for subspace-based sensitivity map estimation. 
\end{abstract}

\begin{IEEEkeywords}
Sensitivity Map Estimation, Multichannel MRI, Parallel Imaging, Structured Low-Rank Matrix Modeling, Autoregression.  
\end{IEEEkeywords}

\section{Introduction}
\label{sec:introduction}
Most modern magnetic resonance imaging (MRI) experiments are performed using   multichannel phased-array receiver coils \cite{roemer1990,hyde1986,carlson1987,hutchinson1988,kelton1989,kwiat1991,ra1993, sodickson1997simultaneous, Pruessmann_1999, Pruessmann_2001,kyriakos2000,griswold2002,lustig2010,zhang2011,blaimer2004smash,ying2010parallel,deshmane2012parallel}.  Assuming $Q$ receiver channels, the measurements are usually modeled using the Fourier transform as
\begin{equation}
d_q(\mathbf{k}_m) = \int c_q(\mathbf{x})\rho(\mathbf{x}) e^{-i2\pi \mathbf{k}_m^T \mathbf{x}} d\mathbf{x} + \eta_{qm}\label{eq:model}
\end{equation}
for $m = 1,\ldots,M$ and $q = 1,\ldots,Q$.  
In this expression, $M$ denotes the number of k-space sampling locations $\mathbf{k}_m\in \mathbb{R}^D$, where $D$ is the dimension of the image;\footnote{For simplicity, we have adopted the use of ``spatial'' notation throughout this paper, which is consistent with typical use cases in 2D or 3D imaging.  In these cases, we would have either $D=2$ with $\mathbf{x} = [x, y]^T$ and $\mathbf{k} = [k_x, k_y]^T$ or $D=3$ with $\mathbf{x} = [x,y,z]^T$ and $\mathbf{k} = [k_x,k_y, k_z]^T$.  However, it should be noted that our approach is also compatible with other scenarios such as spatiotemporal imaging with time-varying sensitivities, which, for example, might use $D=3$ with $\mathbf{x} = [x,y,f]^T$ and $\mathbf{k} = [k_x, k_y, t]^T$.  } $d_q(\mathbf{k}_m)$ denotes the complex-valued k-space data measured from the $q$th receiver channel at the $m$th k-space location; $\rho(\mathbf{x})$ is the complex-valued underlying MR image (reflecting the state of the excited magnetization at the time of data acquisition)  as a function of the spatial position $\mathbf{x}\in \mathbb{R}^D$; $c_q(\mathbf{x})$ is a complex-valued sensitivity map describing the sensitivity of the $q$th receiver channel to excited magnetization at spatial position $\mathbf{x}$; and $\eta_{qm}$ represents complex-valued noise.     We use $\rho_q(\mathbf{x}) \triangleq c_q(\mathbf{x})\rho(\mathbf{x})$ to denote the $q$th sensitivity-weighted image.

This paper is focused on the problem of estimating the sensitivity maps $c_q(\mathbf{x})$ from a set of calibration measurements (i.e., a set of k-space data samples that are sampled at the Nyquist rate). While prior knowledge of the coil sensitivity maps is not always required, there are many tasks that can benefit from estimated sensitivity maps, including coil combination (in which the underlying MRI image $\rho(\mathbf{x})$ is estimated from the multichannel images $\rho_q(\mathbf{x})$ \cite{roemer1990,walsh2000adaptive,bydder2002}) and accelerated image reconstruction (in which the underlying MRI image $\rho(\mathbf{x})$ is estimated from multichannel k-space data $d_q(\mathbf{k}_m)$ that is sampled below the Nyquist rate \cite{sodickson1997simultaneous,Pruessmann_1999,Pruessmann_2001,kyriakos2000,blaimer2004smash,ying2010parallel,deshmane2012parallel}).  

Over the years, many different sensitivity map estimation methods have been proposed \cite{sodickson1997simultaneous,Pruessmann_1999,walsh2000adaptive, keeling2004variational, ying2007joint,  uecker2008,  allison2012accelerated, morrison2007multichannel, uecker2014, gungor2014subspace, she2015, holme2019, bydder2002, peng2022deepsense}.  Among these, subspace-based approaches \cite{morrison2007multichannel, gungor2014subspace, she2015, uecker2014} have proven to be particularly powerful and popular, with the subspace-based ESPIRiT method \cite{uecker2014} rising to become the  most widely-used sensitivity map estimation method in the modern literature because of its simplicity (with few tuning parameters) and excellent performance.

The mathematical principles underlying subspace-based sensitivity map estimation are nontrivial and are not always easy to understand.  In the first part of this work, we present a novel theoretical derivation of subspace-based sensitivity map estimation that is based on a nullspace/linear predictability perspective \cite{haldar2019}, and makes use of concepts from the recent literature on structured low-rank matrix modeling \cite{zhang2011, shin2014, haldar2014low, Haldar_2015,haldar2015autocalibrated, jin2015a, ongie2016, haldar2019,jacob2020}.  This approach is distinct from and complementary to existing theoretical explanations, and we personally find it to be more intuitive than the alternatives.

In the second part of this work, we  introduce several new computational approaches that can substantially reduce the time and memory required for subspace-based sensitivity map estimation. This can be important because existing subspace-based methods can be computationally demanding, with  substantial memory requirements and slow computation times. While slow computations can already be problematic when considering individual datasets, they can be especially limiting in modern machine learning contexts where it may be of interest to calculate sensitivity maps for every element of a database containing hundreds or thousands of images.
 
Our computational contributions include the following:
\begin{itemize}
 \item We introduce a fast FFT-based approach that allows us to rapidly calculate the nullspace vectors associated with a multichannel convolution-structured matrix without  directly constructing the   matrix.  Our approach can be viewed as a multichannel generalization of an approach that was previously developed for single-channel MRI reconstruction \cite{ongie2017fast}.    In addition to being useful for subspace-based sensitivity map estimation, this type of approach may also be useful for the broad class of structured low-rank modeling methods for multichannel MRI image reconstruction \cite{zhang2011, haldar2019,haldar2014low,shin2014,Haldar_2015,haldar2015autocalibrated,jin2015a,ongie2016,jacob2020}. Note that a similar FFT-based approach was proposed for  multichannel MRI reconstruction in Ref.~\cite{zhao2021high}. Compared to that approach, our implementation uses substantially fewer FFTs, which is enabled by making use of structural redundancies.
 \item We leverage the previous observation that using ellipsoidal convolution kernels instead of conventional rectangular kernels can lead to substantial reductions in memory usage and computation without sacrificing accuracy \cite{lobos2022shape}.
\item Subspace-based methods often require constructing a small matrix for every spatial location in the image, where the subspaces of the matrix reflect sensitivity information \cite{uecker2014,walsh2000adaptive}.  We propose an efficient approach that uses vector-space simplifications and FFTs to directly calculate dimension-reduced versions of these matrices.
 \item Noting that sensitivity maps arise from Maxwell's equations and  must be smooth, we propose to estimate the sensitivity maps on a low-resolution spatial grid and interpolate to the desired spatial resolution.
 \item A common step in subspace-based methods is to compute the singular value decomposition (SVD) of a large number of small matrices (i.e., one SVD for each  spatial location)  \cite{uecker2014,walsh2000adaptive}.   We introduce an efficient iterative approach (based on classical power iteration \cite{golub1996}) to calculate many small-scale partial SVDs simultaneously. While this approach is directly useful for subspace-based sensitivity map estimation, we expect that it may also have some utility for a range of locally low-rank modeling and denoising methods that also require calculating partial SVDs over local image patches for every spatial location in the image \cite{trzasko2011a,trzasko2012,manjon2013,veraart2016,cordero-grande2019}.
 \end{itemize}
 Combined together, these algorithmic improvements --- which we collectively call PISCO (Power iteration over simultaneous patches, Interpolation, ellipSoidal kernels, and FFT-based COnvolution) ---  enable a substantial computational acceleration for subspace-based sensitivity map estimation.  
 
 This paper is organized as follows.  Section~\ref{sec:ambiguity} describes fundamental ambiguities in sensitivity map estimation that will be relevant in the sequel.  Section~\ref{sec:null} describes our novel derivation of subspace-based sensitivity map estimation from a linear predictability/structured low-rank perspective.  Section~\ref{sec:5} describes and evaluates the PISCO approaches for fast computations.  Finally, discussion and conclusions are presented in Sec.~\ref{sec:disc}.
 
 \section{Ambiguities in the Inverse Problem}\label{sec:ambiguity}
 
 Before describing our novel theory and methods, we first review inherent ambiguities in sensitivity map estimation.
 
 As already mentioned, we are interested in the estimation of sensitivity maps $c_q(\mathbf{x})$ for $q=1,\ldots,Q$ from a set of Nyquist-sampled calibration measurements.  We assume that these calibration measurements consist of a collection of $M$ k-space samples  from each of the $Q$ channels, for a total of $MQ$ observations. Our goal will be to calculate the values of the sensitivity maps on a discrete voxel grid of $N$ spatial locations $\mathbf{x}_n$ for $n=1,\ldots,N$, resulting in a total of $NQ$ unknowns.  Most practical scenarios will involve substantially more spatial locations than k-space samples (i.e., $NQ \gg MQ$), such that the inverse problem is underdetermined (with fewer measurements than unknowns) unless additional assumptions are made about the characteristics of $c_q(\mathbf{x})$.  To overcome this issue, the literature typically makes the assumption that the sensitivity maps $c_q(\mathbf{x})$ are spatially smooth, which can be justified by physical principles (e.g., Maxwell's equations).
 
 A more pernicious issue is that, while our goal is to estimate the unknown sensitivity maps $c_q(\mathbf{x})$, the image $\rho(\mathbf{x})$ that appears in Eq.~\eqref{eq:model} is also generally an unknown.  This means that Eq.~\eqref{eq:model} can be viewed as a bilinear inverse problem, which is problematic because bilinear inverse problems have well-known scaling ambiguities.  In particular, given one set of estimated sensitivity maps $c_q(\mathbf{x})$ for $q=1,\ldots,Q$ and a corresponding estimated image $\rho(\mathbf{x})$ that together provide a good fit to the measured data, it is straightforward to observe that another equally-good fit to the measured data can be obtained by combining a rescaled version of the sensitivity maps $\alpha(\mathbf{x}) c_q(\mathbf{x})$ for $q=1,\ldots,Q$ together with a rescaled version of the image $\frac{1}{\alpha(\mathbf{x})} \rho(\mathbf{x})$.  The complex-valued scaling function $\alpha(\mathbf{x})$ must be nonzero for every point $\mathbf{x}$ within the field of view (FOV), but is otherwise completely unconstrained, and can have arbitrary spatial variations.  As such, unless additional strong assumptions are made, there will be infinitely many ways of choosing $\alpha(\mathbf{x})$ that all lead to sensitivity map estimates that provide equally-good fits to the measured data.  Practically, this means that both the magnitude and phase of the estimated sensitivity maps are inherently ambiguous, and we can only hope for sensitivity map estimates that are unique up to these scaling ambiguities.  
 
 Because of the inherent ambiguity, it is common for sensitivity map estimation methods to search for one set of estimated maps $c_q(\mathbf{x})$ that fit the data well, and subsequently rescale them with a heuristically-chosen scaling function $\alpha(\mathbf{x})$ that can be designed to endow the scaled sensitivity maps (and the corresponding scaled image) with properties that are deemed to be useful.  Note also that the sensitivity map values $c_q(\mathbf{x})$ will have no impact on measured data at spatial locations where $\rho(\mathbf{x})=0$. As a result, the values of the sensitivity maps are completely unidentifiable at spatial locations outside the support of $\rho(\mathbf{x})$, leading to additional ambiguity.
 
 \section{Nullspace/Linear Predictability Theory for Subspace-Based Sensitivity Map Estimation}\label{sec:null}
 
 In this section, we present a novel derivation of subspace-based sensitivity map estimation from a  nullspace/linear predictability perspective \cite{haldar2019}.  We start by reviewing linear predictability for multichannel MRI before making novel links to sensitivity map estimation.
 
 \subsection{Summary of Linear Predictability for Multichannel MRI}\label{sec:C}
 
 Define the spatial support of $\rho(\mathbf{x})$ as the set of points  $\Omega = \{\mathbf{x} \in \mathbb{R}^D: |\rho(\mathbf{x})|>0\}$.  In what follows, we assume (without loss of generality) that the spatial coordinate system has been normalized such that the support $\Omega$ is completely contained within a hypercube $\Gamma$ with sides of length one, defined as $\Gamma \triangleq \{\mathbf{x} \in \mathbb{R}^D : \|\mathbf{x}\|_\infty < \frac{1}{2}\}$. We will refer to $\Gamma$ as the FOV and use $\Omega^{\complement}\triangleq \Gamma\setminus \Omega$ to denote the complement of $\Omega$ in $\Gamma$. Our FOV assumption allows us to equivalently represent the continuous images $\rho(\mathbf{x})$ and $\rho_q(\mathbf{x})$ using infinite Fourier series representations as
 \begin{equation}
  \rho(\mathbf{x}) = \mathbbm{1}_\Gamma(\mathbf{x})\sum_{\mathbf{n} \in \mathbb{Z}^D} s[\mathbf{n}] e^{i 2\pi \mathbf{n}^T \mathbf{x}}
 \end{equation}
 and
 \begin{equation}
  \rho_q(\mathbf{x}) = \mathbbm{1}_\Gamma(\mathbf{x})\sum_{\mathbf{n} \in \mathbb{Z}^D} s_q[\mathbf{n}] e^{i 2\pi \mathbf{n}^T \mathbf{x}}
 \end{equation}
 for $q=1,\ldots,Q$, where $s[\mathbf{n}]$ and $s_q[\mathbf{n}]$ respectively represent samples of the Fourier transforms of $\rho(\mathbf{x})$ and $\rho_q(\mathbf{x})$ on the rectilinear Nyquist grid (taking $\Delta k=1$ along each dimension, as enabled by our assumptions about $\Gamma$), and $\mathbbm{1}_\Gamma(\mathbf{x})$ is the indicator function for the FOV:
 \begin{equation}
  \mathbbm{1}_\Gamma(\mathbf{x}) = \left\{ \begin{array}{ll} 1, &\mathbf{x} \in \Gamma \\ 0, & \mathrm{else}. \end{array}\right.
 \end{equation}
 
 A basic assumption of the linear predictability formalism \cite{haldar2019} is that  $\{s_q[\mathbf{n}]\}_{q=1}^Q$ will be autoregressive, satisfying multiple multichannel shift-invariant linear predictability relationships \cite{Haldar_2015, haldar2019,haldar2015autocalibrated,shin2014,jin2015a,jacob2020, lustig2010, uecker2014, morrison2007multichannel, she2015, gungor2014subspace, griswold2002,zhang2011,haldar2014low,cheung1990}. Specifically, there should exist multiple distinct $Q$-channel finite impulse response (FIR) filters  $h_r[\mathbf{n},q]$ that each satisfy 
 \begin{equation}
  \sum_{q=1}^Q \sum_{\mathbf{m} \in \Lambda} h_r[\mathbf{m},q] s_q[\mathbf{n}-\mathbf{m}] \approx 0 \text{ for } \forall \mathbf{n} \in \mathbb{Z}^D\label{eq:annihil}
 \end{equation}
 for $r=1,\ldots,R$, where $R$ is the number of distinct filters and  $\Lambda \subset \mathbb{Z}^D$ is a finite index set describing the support of each FIR filter.

There are many different situations for which the existence of such filters $h_r[\mathbf{n},q]$ has been theoretically proven \cite{haldar2019, jacob2020}. We review a few of these below:\footnote{Shift-invariant linear predictability relationships can also be proven for images that possess smooth phase \cite{haldar2014low, haldar2019, huang2009} and sparsity characteristics \cite{haldar2019, liang1989, liang1989a, jin2015a, ongie2016, jacob2020}, although these relationships are not directly relevant for sensitivity map estimation and we do not discuss them further.} 

 \subsubsection{\bf Limited Image Support  \cite{haldar2014low,  cheung1990, haldar2019}}  In many situations, the actual spatial support $\Omega$ of the original image $\rho(\mathbf{x})$ will be smaller than the FOV $\Gamma$, such that the complement of $\Omega$ in $\Gamma$ (denoted  by $\Omega^\complement \triangleq \Gamma  \setminus \Omega$), is a measurable nonempty set.  In such cases, it is possible to define nonzero functions $f(\mathbf{x})$ that have supports fully-contained within $\Omega^\complement$.  Since $\rho(\mathbf{x})$ and $f(\mathbf{x})$ have nonoverlapping support, we must have $f(\mathbf{x})\rho(\mathbf{x}) = 0$ for $\forall \mathbf{x} \in \Gamma$.  If we let $\tilde{f}[\mathbf{n}]$ denote the Nyquist-sampled Fourier series representation of $f(\mathbf{x})$, then applying the Fourier convolution theorem to  the Fourier transform of $f(\mathbf{x})\rho(\mathbf{x})$ yields that
 \begin{equation}
  \sum_{\mathbf{m} \in \mathbb{Z}^D} \tilde{f}[\mathbf{m}] s[\mathbf{n}-\mathbf{m}] = 0 \text{ for } \forall \mathbf{n} \in \mathbb{Z}^D.
 \end{equation}
 If $\Omega^\complement$ is large enough, then it becomes possible to choose relatively smooth functions $f(\mathbf{x})$ that satisfy the conditions specified above.   If we assume that these functions are smooth enough that they satisfy $\tilde{f}[\mathbf{n}] \approx 0$ for $\forall \mathbf{n} \notin \Lambda$ (i.e., the functions $f(\mathbf{x})$ are approximately bandlimited to the frequency range defined by $\Lambda$),\footnote{While we can often obtain very accurate approximations of this form, this must always be an approximation, and will never be exact.  Specifically, since $f(\mathbf{x})$ is assumed to have finite support, then $\tilde{f}[\mathbf{n}]$ cannot have exactly-limited support by the Fourier uncertainty principle \cite{bracewell2000}. While this paper describes this approximation  informally to simplify the exposition,  more formal treatments exist in the literature \cite{haldar2019}. } then we can approximate $\tilde{f}[\mathbf{n}]$ as an FIR filter, and obtain
 \begin{equation} \label{eq:ann_neigh}
  \sum_{\mathbf{m} \in \Lambda} \tilde{f}[\mathbf{m}] s[\mathbf{n}-\mathbf{m}] \approx 0 \text{ for } \forall \mathbf{n} \in \mathbb{Z}^D.
 \end{equation}
Note also that if we have $f(\mathbf{x})\rho(\mathbf{x}) \approx 0$ for $\forall \mathbf{x} \in \Gamma$, then the relationship between $\rho(\mathbf{x})$ and $\rho_q(\mathbf{x})$ implies we must also have that $f(\mathbf{x})\rho_q(\mathbf{x}) \approx 0$ for $\forall \mathbf{x} \in \Gamma$ for $q=1,\ldots,Q$.  This implies that the multichannel FIR filter 
 \begin{equation}
  h_r[\mathbf{n},q] = \left\{ \begin{array}{cl} \alpha_q \tilde{f}[\mathbf{n}], & \mathbf{n} \in \Lambda \\ 0, & \mathrm{else}\end{array}\right.\label{eq:f1}
 \end{equation}
 will satisfy Eq.~\eqref{eq:annihil} for any choices of $\alpha_q \in \mathbb{C}$ for $q=1,\ldots,Q$.  It is easy to obtain many distinct $h_r[\mathbf{n},q]$ filters from this relationship because of the flexibility offered by  choosing different functions $f(\mathbf{x})$ and different coefficients $\alpha_q$.  Using similar arguments, additional channel-specific filters can also be obtained in common situations where some sensitivity-weighted images $\rho_q(\mathbf{x})$ have smaller support than the underlying image $\rho(\mathbf{x})$.
\subsubsection{\bf Smooth Sensitivity Maps \cite{haldar2019,jacob2020,morrison2007multichannel,she2015}}  As already mentioned, sensitivity maps are often spatially smooth.  Linear prediction relationships can be derived if we assume the $c_q(\mathbf{x})$ are so smooth that they are approximately bandlimited to the index set $\Lambda$, such that
\begin{equation}
 c_q(\mathbf{x}) \approx \mathbbm{1}_\Gamma(\mathbf{x})\sum_{\mathbf{n}\in \Lambda} \tilde{c}_q[\mathbf{n}] e^{i2\pi \mathbf{n}^T\mathbf{x}}
\end{equation}
for appropriate Fourier coefficients $\tilde{c}_q[\mathbf{n}]$.  To see why, observe that $c_i(\mathbf{x}) \rho_j(\mathbf{x}) - c_j(\mathbf{x}) \rho_i(\mathbf{x}) = 0$ for $\forall \mathbf{x} \in \Gamma$ for any choices of $i$ and $j$ (sometimes called the ``cross-relation'' in the multichannel deconvolution literature \cite{gurelli1995,xu1995,harikumar1999}).  Applying bandlimitedness and the Fourier convolution theorem to the cross-relation leads to
\begin{equation}
\begin{split}
 \sum_{\mathbf{m} \in \Lambda} &\tilde{c}_i[\mathbf{m}] s_j[\mathbf{n}-\mathbf{m}] \\ &- \sum_{\mathbf{m}\in\Lambda} \tilde{c}_j[\mathbf{m}] s_i[\mathbf{n}-\mathbf{m}] \approx 0 \text{ for } \forall \mathbf{n} \in \mathbb{Z}^D.\\
 \end{split}
\end{equation}
This implies that the multichannel FIR filter
\begin{equation}
 h_r[\mathbf{n},q] = \left\{\begin{array}{cl} 
 -\tilde{c}_j[\mathbf{n}], & \mathbf{n} \in \Lambda, q=i \\ 
 \tilde{c}_i[\mathbf{n}], & \mathbf{n} \in \Lambda, q=j \\
 0, &\mathrm{else}\end{array} \right.\label{eq:f2}
\end{equation}
will satisfy Eq.~\eqref{eq:annihil} for any choices of $i$ and $j$ with $i\neq j$.  It is easy to obtain many distinct $h_r[\mathbf{n},q]$ filters using different choices of $i$ and $j$.

\subsubsection{\bf Compressible Receiver Arrays \cite{buehrer2007, huang2008, kim2021}} It is frequently observed that the different channels in a multichannel receiver array possess linear dependencies, such that there  exist multiple distinct choices of linear combination weights $w_q$ that will each satisfy
\begin{equation}
 \sum_{q=1}^Q w_q \rho_q(\mathbf{x}) \approx 0 \text{ for } \forall \mathbf{x} \in \Gamma.\label{eq:gamma}
\end{equation}
This observation is often used for dimensionality reduction (``coil compression'') \cite{buehrer2007, huang2008, kim2021}.  However, this observation also implies that a relationship in the form of Eq.~\eqref{eq:annihil} will hold if $h_r[\mathbf{n},q]$ is chosen as
\begin{equation}
 h_r[\mathbf{n},q] = \left\{\begin{array}{ll} w_q, & \mathbf{n}=\mathbf{0} \\ 0, & \text{else}. \end{array} \right.\label{eq:f3}
\end{equation}
Since there often exist multiple different choices of $w_q$ that satisfy Eq.~\eqref{eq:gamma}, there can also be many distinct filters $h_r[\mathbf{n},q]$ that arise from this kind of relationship.

The preceding arguments all support the existence of multiple filters $h_r[\mathbf{n},q]$ that  satisfy Eq.~\eqref{eq:annihil}.  If we were given full information about the support $\Omega$ and the sensitivity maps $c_q(\mathbf{x})$, then we would be able to calculate filters $h_r[\mathbf{n},q]$ directly from Eqs.~\eqref{eq:f1}, \eqref{eq:f2}, and \eqref{eq:f3}.  However, in practice, we do not have information about $\Omega$ or $c_q(\mathbf{x})$ (these are quantities that we are interested in estimating!), and we must instead resort to other methods to obtain the FIR filters $h_r[\mathbf{n},q]$. 
 
An important observation from the literature on structured low-rank matrix modeling  \cite{haldar2019,jacob2020} is that the convolution appearing in Eq.~\eqref{eq:annihil} can be represented in matrix form as $\mathbf{C}\mathbf{h}_r \approx \mathbf{0}$, where the matrix $\mathbf{C}$ is given by
\begin{equation}
\mathbf{C} \triangleq \begin{bmatrix} \mathbf{C}_1 & \mathbf{C}_2 & \cdots & \mathbf{C}_Q  \end{bmatrix};\label{eq:C}
\end{equation}
the matrix $\mathbf{C}_q \in \mathbb{C}^{P\times |\Lambda|}$ is a convolution-structured matrix (i.e., a Hankel or Toeplitz matrix) formed from Fourier samples $s_q[\mathbf{n}]$, with each row of $\mathbf{C}_q$ corresponding to one value of $\mathbf{n}$ from Eq.~\eqref{eq:annihil}; $P$ represents the number of different $\mathbf{n}$ values that are used in the construction of each $\mathbf{C}_q$ matrix; and $\mathbf{h}_r \in \mathbb{C}^{Q|\Lambda|}$ is the vector of $h_r[\mathbf{n},q]$ samples.  See Ref.~\cite{Haldar_2015} for a detailed description of the matrix construction we use.

This matrix representation shows that all filters $h_r[\mathbf{n},q]$ that satisfy Eq.~\eqref{eq:annihil} must be approximate nullspace vectors of an appropriately-constructed $\mathbf{C}$ matrix.  Importantly, it is possible to directly form such a $\mathbf{C}$ matrix from Nyquist-sampled calibration data, and its approximate nullspace can be identified using the SVD.  It is therefore possible to learn a set of $R$ linearly independent filters $h_r[\mathbf{n},q]$ (with $R$ equal to the dimension of the approximate nullspace of $\mathbf{C}$) in a data-driven way  \cite{zhang2011, haldar2015autocalibrated}.  The full set of approximate nullspace vectors is expected to form a basis for the set of all possible $h_r[\mathbf{n},q]$ filters that are consistent with Eq.~\eqref{eq:annihil} and bandlimited to $\Lambda$.

\subsection{Linear Predictability and Sensitivity Map Estimation}\label{sec:HG}
In what follows, we assume that we have access to $R$ filters $h_r[\mathbf{n},q]$ that each satisfy Eq.~\eqref{eq:annihil}, which can, e.g., be obtained from the nullspace of a calibration matrix as described above.

Observe that the linear prediction relationship from Eq.~\eqref{eq:annihil} can be rewritten in the image domain as
\begin{equation}
 \sum_{q=1}^Q h_r(\mathbf{x},q) \rho_q(\mathbf{x}) \approx 0 \text{ for } \forall \mathbf{x} \in \Gamma,\label{eq:annihil2}
\end{equation}
where
\begin{equation}
 h_r(\mathbf{x},q) \triangleq \sum_{\mathbf{n}\in\Lambda} h_r[\mathbf{n},q] e^{i2\pi \mathbf{n}^T\mathbf{x}}.
\end{equation}
Using the relationship between $\rho(\mathbf{x})$ and $\rho_q(\mathbf{x})$, Eq.~\eqref{eq:annihil2} can be further simplified to
\begin{equation}
   \rho(\mathbf{x}) \sum_{q=1}^Q h_r(\mathbf{x},q) c_q(\mathbf{x}) \approx 0 \text{ for } \forall \mathbf{x} \in \Gamma.\label{eq:filt1}
\end{equation}
Since $|\rho(\mathbf{x})| > 0$ for $\forall\mathbf{x} \in \Omega$ (by the definition of the image support set $\Omega$), this implies that  the summation
\begin{equation}
   \sum_{q=1}^Q h_r(\mathbf{x},q) c_q(\mathbf{x}) \approx 0 \text{ for } \forall \mathbf{x} \in \Omega, \label{eq:filt}
\end{equation}
demonstrating that the sensitivity maps $c_q(\mathbf{x})$ and the spatial-domain representation of the filters $h_r(\mathbf{x},q)$ have strong dependencies that must be satisfied for all $\mathbf{x}\in\Omega$.
On the other hand, the summation  in Eq.~\eqref{eq:filt} could result in arbitrary nonzero values for $\mathbf{x} \in \Omega^\complement$, since the fact that $\rho(\mathbf{x})=0$ at these spatial locations will still ensure that Eq.~\eqref{eq:filt1} is satisfied regardless of the properties of $h_r(\mathbf{x},q)$ and $c_q(\mathbf{x})$.

Since we have $R$ filters $h_r[\mathbf{n},q]$ that each satisfy Eq.~\eqref{eq:annihil}, the relationship from Eq.~\eqref{eq:filt} can be expressed simultaneously for all $R$ filters  in matrix form as
\begin{equation}
 \mathbf{H}(\mathbf{x}) \mathbf{c}(\mathbf{x}) \approx \mathbf{0} \text{ for } \forall \mathbf{x} \in \Omega,
\end{equation}
 where $\mathbf{H}(\mathbf{x}) \in \mathbb{C}^{R \times Q}$ has entries  $[\mathbf{H}(\mathbf{x})]_{rq} = h_r(\mathbf{x},q)$, and $\mathbf{c}(\mathbf{x}) \in \mathbb{C}^Q$ is the vector with the $q$th entry equal to $c_q(\mathbf{x})$. Importantly, this demonstrates that the sensitivity map values at location $\mathbf{x} \in \Omega$ will be one of the approximate nullspace vectors of $\mathbf{H}(\mathbf{x})$.  Ideally, there is only one approximate nullspace vector at the the spatial location $\mathbf{x} \in \Omega$.  In this case, because singular vectors associated with distinct singular values are always unique up to scaling ambiguities, we would able to use the unique approximate nullspace vector to obtain unique sensitivity map values $c_q(\mathbf{x})$  up to the same scaling ambiguities that are inherent to sensitivity map estimation (see previous discussion in Sec.~\ref{sec:ambiguity}).
 
 The ability to estimate  unique sensitivity maps $c_q(\mathbf{x})$ (up to scaling ambiguities) for $\mathbf{x} \in \Omega$  requires that the matrix $\mathbf{H}(\mathbf{x})$ is approximately rank $Q-1$.  A necessary condition for this to be true is that $R \geq Q-1$, although this condition is not sufficient (and it is easy to find counter examples where $R\geq Q-1$ but $\mathbf{H}(\mathbf{x})$ does not have the desired characteristics).  In practice, we have observed empirically that if we use the entire collection of filters from the approximate nullspace of $\mathbf{C}$,\footnote{Our experience is based on applying a heuristic thresholding rule to the SVD to determine the approximate nullspace.  Specifically, if $\sigma_n$ denotes the $n$th singular value of $\mathbf{C}$ (with singular values indexed in order from largest to smallest), then  the $n$th right singular vector is included in the approximate nullspace whenever  $\sigma_n < 0.05 \sigma_1$.  The $0.05\sigma_1$ threshold was determined heuristically and is unlikely to be optimal, although worked well for the datasets considered in this work.}  then the $\mathbf{H}(\mathbf{x})$ matrices  always have an approximate rank of $Q-1$ for $\mathbf{x} \in \Omega$ under the assumptions and scenarios considered in this paper.\footnote{Different rank characteristics may be observed under different assumptions.  For example, if the image support $\Omega$ were larger than the FOV $\Gamma$ such that aliasing occurs, then we empirically observe multiple sets of nullspace vectors at the spatial locations where the image support aliases onto itself.  This  is expected since such aliased voxels will be a linear combination of multiple sets of sensitivity-weighted images, each with distinct sensitivity maps.  This enables the computation of multiple sensitivity maps, similar to what has been described previously in work on ESPIRiT  \cite{uecker2014}, although we do not belabor this point as this scenario is not our primary interest.}  Interestingly, when $\mathbf{x} \in \Omega^\complement$ (i.e., the situation where the sensitivity maps are unidentifiable) and we use the entire collection of filters from the approximate nullspace of $\mathbf{C}$, we empirically observe that the matrix $\mathbf{H}(\mathbf{x})$ always has rank $Q$ rather than $Q-1$, which provides a simple mechanism for identifying both $\Omega$ and $\Omega^\complement$.

Empirical demonstrations of the  nullspace characteristics of $\mathbf{C}$ and $\mathbf{H}(\mathbf{x})$  are presented in the next subsection.

\subsection{Empirical Demonstration of Nullspace Characteristics}\label{sec:testdata}
\begin{figure}[t]
\centering 
 \begin{minipage}{0.31\linewidth}
 \renewcommand\thesubfigure{(a) Brain MPRAGE}
\subfigure[]{\includegraphics[width=0.8425\linewidth]{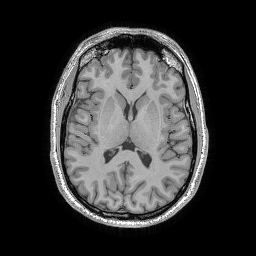}} 
 \end{minipage}
 \hspace{-6mm}
 \begin{minipage}{0.33\linewidth}
 \renewcommand\thesubfigure{(b) Brain TSE}
\subfigure[]{\includegraphics[width=1.05\linewidth]{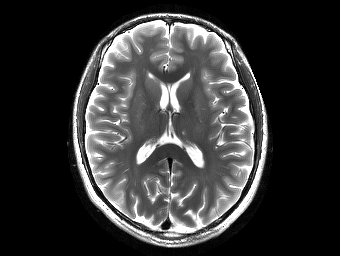}}
\end{minipage}
 \hspace{-0.2mm}
 \begin{minipage}{0.3\linewidth}
 \renewcommand\thesubfigure{(c) Knee TSE}
\subfigure[]{\includegraphics[width=1\linewidth]{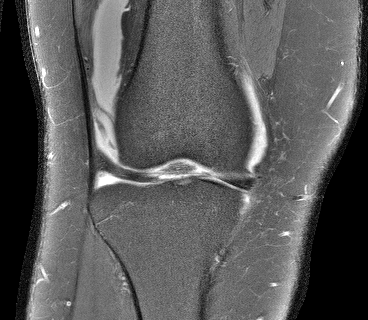}}
\end{minipage}
\caption{Depiction of the three datasets we use for illustration and validation. }
\label{fig:images}
\end{figure}

Throughout this paper, we will use the following three multichannel MRI datasets to illustrate theoretical principles and to validate our proposed methods:
\begin{itemize}
 \item {\bf Brain MPRAGE.} We used an MPRAGE sequence on a 3T scanner to acquire a T1-weighted human brain image using a 32-channel receiver array, as depicted in Fig.~\ref{fig:images}(a). 
 \item {\bf Brain TSE.} We used a turbo spin-echo sequence on a 3T scanner to acquire a T2-weighted human brain image using a 32-channel receiver array, as depicted in Fig.~\ref{fig:images}(b).
 \item {\bf Knee TSE.} We used a 15-channel knee dataset from the fastMRI database  \cite{zbontar2019}, as depicted in Fig.~\ref{fig:images}(c).  This image was acquired using a fat-suppressed turbo spin-echo sequence on a 3T scanner, producing a proton-density weighted image. 
\end{itemize}

\begin{figure}[t]
 \centering 
 \hspace{-4mm}
 \begin{minipage}{0.3\linewidth}
 \renewcommand\thesubfigure{(a) Brain MPRAGE}
\subfigure[]{\includegraphics[width=1.1\linewidth]{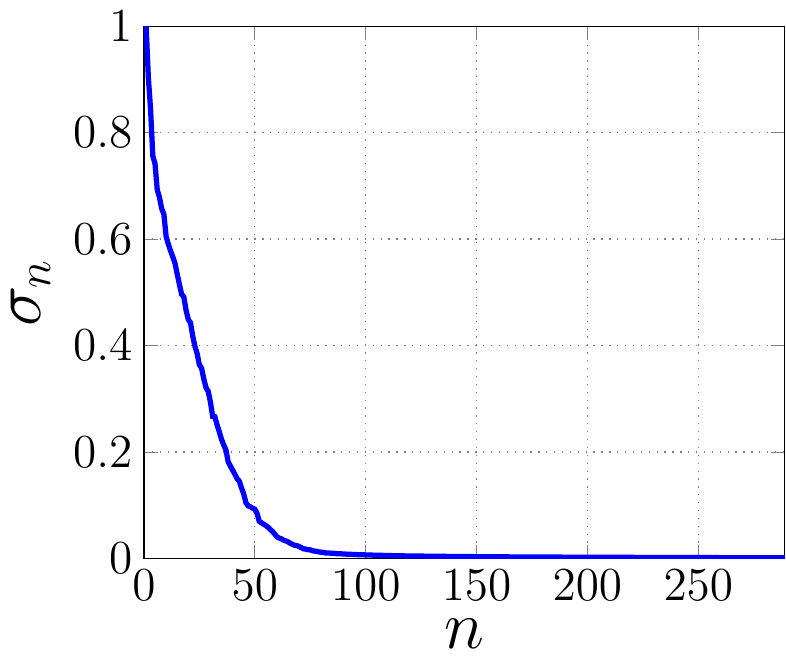}} 
 \end{minipage}
 \hspace{1mm}
 \begin{minipage}{0.3\linewidth}
 \renewcommand\thesubfigure{(b) Brain TSE}
\subfigure[]{\includegraphics[width=1.1\linewidth]{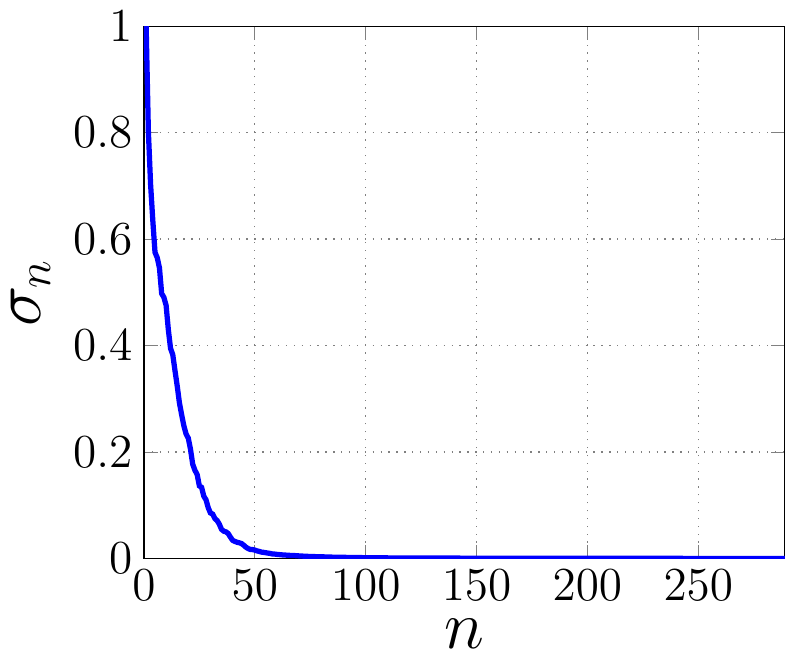}}
\end{minipage}
 \hspace{1mm}
 \begin{minipage}{0.3\linewidth}
 \renewcommand\thesubfigure{(c) Knee TSE}
\subfigure[]{\includegraphics[width=1.1\linewidth]{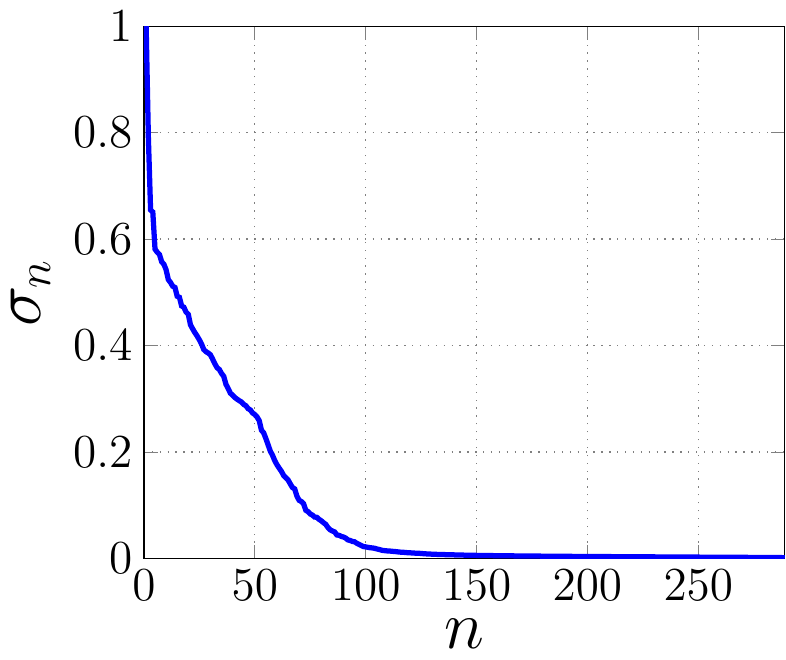}}
\end{minipage}
 \caption{Normalized singular value curves for the $\mathbf{C}$ matrices corresponding to the three datasets from Fig.~\ref{fig:images}. }
 \label{fig:plots_C_sv}
 \end{figure}

To illustrate the low-rank characteristics of the $\mathbf{C}$ matrix as described in Sec.~\ref{sec:C}, Fig.~\ref{fig:plots_C_sv} plots the singular values of $\mathbf{C}$ for all three datasets.  In each case, the $\mathbf{C}$ matrix was calculated using the central 24$\times$24 k-space samples from the Nyquist grid and assuming a $7\times 7$ rectangular FIR filter support (i.e., $\Lambda = \{ \mathbf{n} \in \mathbb{Z}^2: \|\mathbf{n}\|_\infty \leq 3\}$). As expected,  all three $\mathbf{C}$ matrices are approximately low-rank with a substantial approximate nullspace.

\begin{figure}[t]
 \centering 
\includegraphics[width=0.5\textwidth]{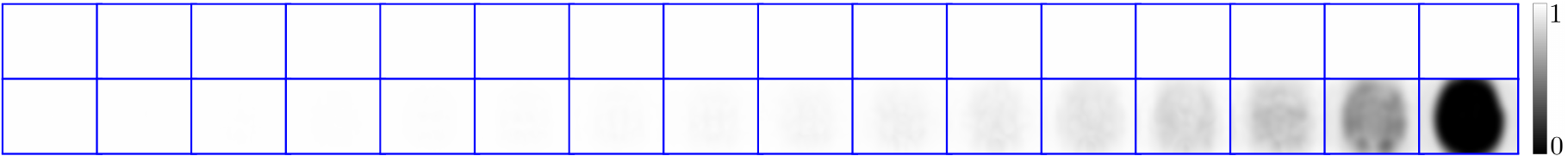}
 \caption{Spatial maps depicting the 32  singular values of $\mathbf{H}(\mathbf{x})$ as a function of $\mathbf{x}$ for the Brain TSE dataset. We show maps of $\frac{1}{|\Lambda|} \sigma_n^2$ (i.e., the squared singular values normalized by $|\Lambda|$) in sequential order from left to right and top to bottom.  }
 \label{fig:sv_H}
 \end{figure}

 To illustrate the characteristics of the $\mathbf{H}(\mathbf{x})$ matrix described in Sec.~\ref{sec:HG}, Fig.~\ref{fig:sv_H} shows spatial maps of the $Q=32$ squared singular values of $\mathbf{H}(\mathbf{x})$ for the Brain TSE dataset.  As expected, the smallest singular value is approximately zero within the support of the image (i.e., $\mathbf{H}(\mathbf{x})$ is approximately rank $= Q-1$ for $\mathbf{x} \in \Omega$), while $\mathbf{H}(\mathbf{x})$ has full rank (i.e., rank $=Q$) outside the image support. 
 
\subsection{Naive Algorithm for Nullspace-Based Sensitivity Map Estimation} \label{sec:alg_nullspace}
Based on the theoretical framework described in the previous subsections, a naive nullspace-based algorithm for calulating sensitivity maps would proceed as follows:
\begin{enumerate}
 \item Choose an FIR filter support set $\Lambda$, and construct the corresponding $\mathbf{C}$ matrix from calibration data. 
 \item Calculate a set of $R$ approximate nullspace vectors of the $\mathbf{C}$ matrix using the SVD.
 \item Using the nullspace vectors from the previous step, calculate the $\mathbf{H}(\mathbf{x})$ matrices for each spatial position $\mathbf{x}$.
 \item For each spatial location $\mathbf{x}$, calculate the SVD of $\mathbf{H}(\mathbf{x})$  and set $c_q(\mathbf{x})$ equal to the $q$th entry of the right singular vector associated with the smallest singular value. 
 \item (Optional) If desired, choose a scaling function $\alpha(\mathbf{x})$ that has useful properties and use it to normalize the sensitivity maps.
  \item (Optional) If desired, obtain a binary mask for the image support $\Omega$ by applying thresholding to the spatial map of the smallest singular value of $\mathbf{H}(\mathbf{x})$.
\end{enumerate}
As we will demonstrate in the sequel, this naive algorithm works well although can require substantial computational complexity and memory usage, while a modified algorithm that uses our proposed PISCO approach leads to substantial computational accelerations and memory savings.

 \begin{figure*}[t]
 \centering 
{\setlength{\fboxsep}{0 mm}\fcolorbox{white}{white}{\includegraphics[width=1\textwidth]{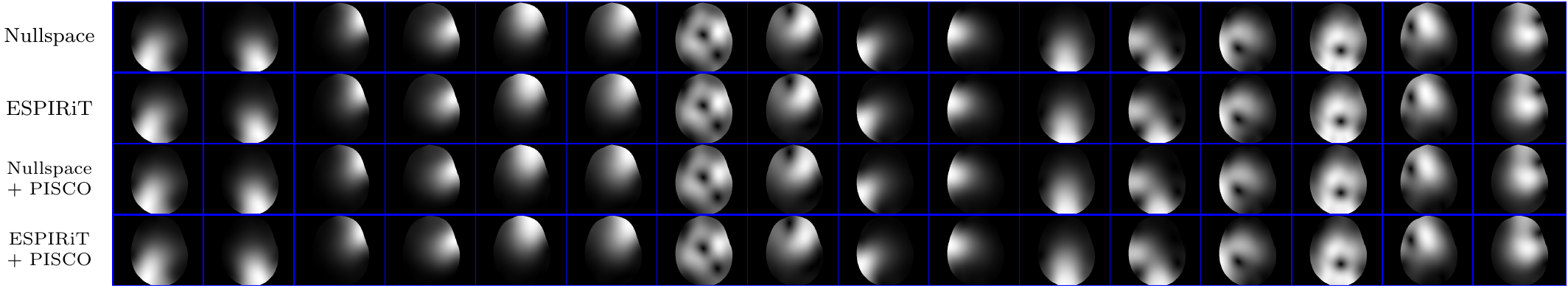}}}
 \caption{ Representative sensitivity maps (16 out of 32 channels from the Brain TSE dataset) obtained from the nullspace-based approach, ESPIRiT, the nullspace-based approach + PISCO, and ESPIRiT + PISCO.  All results used the same calibration data (24$\times$24 central k-space samples) and the same $\Lambda$ (7$\times$7 rectangular).  }
 \label{fig:ESP_null_maps}
 \end{figure*}

\begin{figure}[t]
 \centering 
 \begin{minipage}{0.55\linewidth} 
\subfigure{\includegraphics[width=1.25in]{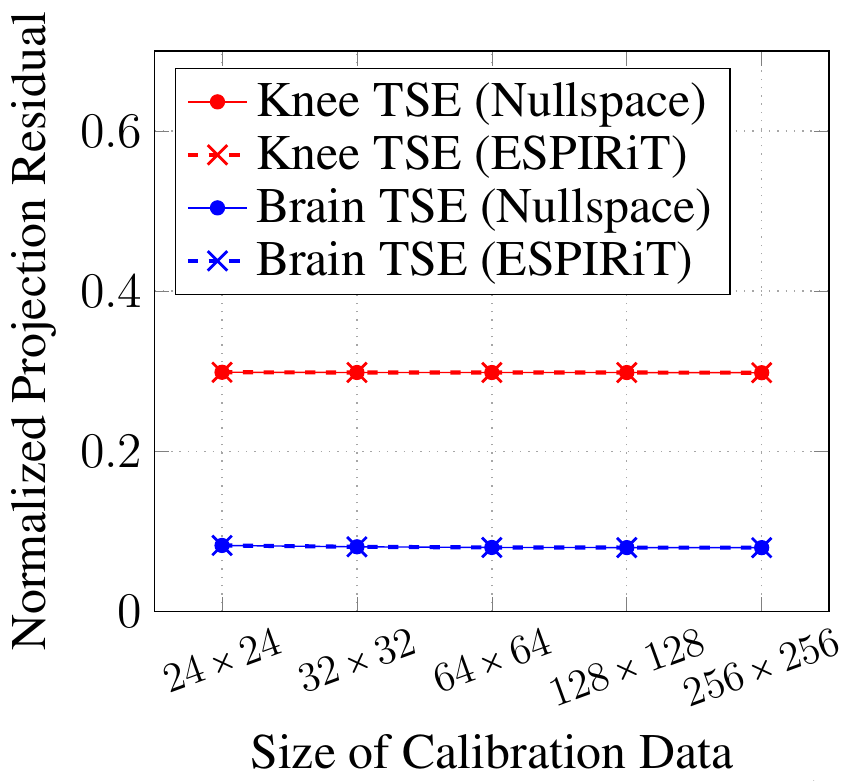}} 
 \end{minipage}
 \caption{ Quantitative performance evaluation (as measured with the normalized projection residual) of the nullspace-based approach and ESPIRiT.}
 \label{fig:ESP_null}
 \end{figure}

\subsection{Comparisons of the Nullspace Algorithm and ESPIRiT}\label{sec:comp_theory}

Before moving on to describe the rest of our novel PISCO computational acceleration techniques, we will first spend a  moment to describe how the nullspace-based approach compares to the popular ESPIRiT approach for sensitivity map estimation \cite{uecker2014}, which is also subspace-based. While the two approaches are derived using very different arguments, the resulting approaches end up having quite similar structure.  Specifically, in our nullspace-based approach, we use the nullspace vectors of the $\mathbf{C}$ matrix to identify matrices $\mathbf{H}(\mathbf{x})$ that satisfy $\mathbf{H}(\mathbf{x}) \mathbf{c}(\mathbf{x}) \approx \mathbf{0}$, where $\mathbf{c}(\mathbf{x}) \in \mathbb{C}^Q$ is the vector of $c_q(\mathbf{x})$ sensitivity map values.  This causes us to estimate sensitivity maps using the approximate nullspace vector associated with the smallest singular value of $\mathbf{H}(\mathbf{x})$.  In contrast, ESPIRiT uses the rowspace of the $\mathbf{C}$ matrix to construct matrices $\mathbf{B}(\mathbf{x}) \in \mathbb{C}^{Q\times Q}$ that are expected to satisfy $\mathbf{B}(\mathbf{x}) \mathbf{c}(\mathbf{x}) \approx \mathbf{c}(\mathbf{x})$.  As a result, the sensitivity maps in ESPIRiT can be estimated using the eigenvector associated with the largest eigenvalue of $\mathbf{B}(\mathbf{x})$.   Interestingly, it can be derived (with a substantial amount of tedious-but-straightforward derivations that are too involved to reproduce here) that the $\mathbf{B}(\mathbf{x})$ matrix used by ESPIRiT will satisfy
\begin{equation}\label{eq:esp_null}
 \mathbf{B}(\mathbf{x}) = \mathbf{I}_Q - \frac{1}{|\Lambda|} (\mathbf{H}(\mathbf{x}))^H \mathbf{H}(\mathbf{x}),
\end{equation}
where $\mathbf{I}_Q$ denotes the $Q\times Q$ identity matrix.  This implies that the results of the nullspace-based approach and ESPIRiT will be formally equivalent if the same parameters are used for both methods.
 
Our empirical results confirm that ESPIRiT and the nullspace-based approach produce identical sensitivity map estimates (up to numerical errors resulting from finite precision arithmetic).  To illustrate this, we used  in-house MATLAB implementations of both approaches\footnote{Note that the naive nullspace algorithm described in Sec.~\ref{sec:alg_nullspace} can require an excessive amount of memory that exceeds MATLAB's limits on the maximum array size.  As such, the results we present as the nullspace-based approach in this section actually use one component of the PISCO approach to bypass this issue, namely the efficient calculation of a dimension-reduced matrix $\mathbf{G}(\mathbf{x})$ instead of the matrix $\mathbf{H}(\mathbf{x})$ as will be described in Sec.~\ref{sec:G_fast}.}  to estimate sensitivity maps for all three datasets from Fig.~\ref{fig:images} using different amounts of calibration data (ranging from small 24$\times$24 calibration regions to larger 256$\times$256 regions, always centered at the k-space origin).  As a precursor to our description of PISCO in the sequel, we also obtained results from the nullspace-based approach combined with PISCO and ESPIRiT combined with PISCO.  Representative qualitative results are illustrated in Fig.~\ref{fig:ESP_null_maps}, which demonstrate negligible visual differences between all four approaches.  

To be more quantitative, we have also computed a ``normalized projection residual''  metric to quantify how well the estimated sensitivity maps match the fully-sampled data.  Specifically, we first take the fully-sampled multichannel data and project it onto the subspace spanned by the sensitivity maps \cite{uecker2014}.\footnote{It was necessary to use a specialized error metric because the inherent scaling ambiguities described in Sec.~\ref{sec:ambiguity} confound traditional error metrics.}  If sensitivity map estimation worked well, then the projection onto the sensitivity maps should capture all of the true signal, with only noise remaining in the residual after projection.  Our metric is obtained as the $\ell_2$-norm of the projection residual divided by the $\ell_2$-norm of the original multichannel data, with smaller values interpreted as  better.  

Figure~\ref{fig:ESP_null}  plots this metric for the Knee TSE dataset (with a fixed 5$\times$5 rectangular FIR filter support $\Lambda$) and the Brain TSE dataset (with a fixed 7$\times$7 rectangular FIR filter support $\Lambda$) for different calibration sizes, while Table~\ref{tab:nrmse} shows numerical results  for all three datasets using 32$\times$32 calibration data.  As before, the differences between ESPIRiT and the nullspace-based approach are negligible. (This table also shows PISCO results which are also quite similar, although we defer  a detailed discussion until after we have described PISCO completely).

 \begin{table}[t]
\caption{Normalized Projection Residual with 32$\times$32 calibration data.}
\centering
{\setlength{\tabcolsep}{2pt}
\begin{tabular}{|c|c|c|c|c|}
\hline
\bfseries Dataset & \bfseries ESPIRiT& \bfseries ESPIRiT &\bfseries  Nullspace & \bfseries Nullspace \\
\bfseries  & \bfseries & \bfseries + PISCO &\bfseries  & \bfseries + PISCO 
\\ 
\hline
Brain MPRAGE &0.136&0.141&0.136&0.141\\\hline
Brain TSE&0.081&0.087&0.081&0.087\\\hline
Knee TSE&0.297&0.298&0.297&0.298\\\hline
\end{tabular}
}
\label{tab:nrmse}
\end{table}

Overall, these results confirm that our nullspace-based approach produces sensitivity maps that are numerically equivalent to the sensitivity maps produced by ESPIRiT.  Although this is expected theoretically, it is  notable given that the two approaches were derived from distinct perspectives.   This concludes  the first major contribution of this paper.  

It should be noted that both ESPIRiT and the nullspace-based approach can require substantial computation times (on the order of minutes) and memory (on the order of gigabytes), as reflected in the computation times and memory usage reported in Tables~\ref{tab:time} and \ref{tab:mem}, respectively.\footnote{Note that, as described previously, the naive nullspace-based approach was not directly implemented because we would exceed MATLAB's limits on the maximum array size, so we do not report computation times and only report projected values (denoted with an asterisk) for memory usage. } The remainder of this paper will focus on another major contribution:  the PISCO computational approaches that are applicable to a range of subspaced-based approaches, including the nullspace-based approach and ESPIRiT.   As can already be seen from Tables~\ref{tab:time} and \ref{tab:mem}, the use of PISCO can substantially improve computation time and memory usage relative to baseline implementations, and both the nullspace-based approach and ESPIRiT have nearly identical computational complexity when PISCO is used.

 \begin{table}[t]
\caption{Computation Time (secs.) with 32$\times$32 calibration data.}
\centering
{\setlength{\tabcolsep}{2pt}
\begin{tabular}{|c|c|c|c|c|}
\hline
\bfseries Dataset & \bfseries ESPIRiT& \bfseries ESPIRiT &\bfseries  Nullspace & \bfseries Nullspace \\
\bfseries  & \bfseries & \bfseries + PISCO &\bfseries  & \bfseries + PISCO 
\\ 
\hline
Brain MPRAGE &41.1&0.8&n/a&0.8\\\hline
Brain TSE&97.4&0.9&n/a&0.9\\\hline
Knee TSE&64.0&0.3&n/a&0.3\\\hline
\end{tabular}
}
\label{tab:time}
\end{table}

\begin{table}[t]
\caption{Memory Usage (GB) with 32$\times$32 calibration data. }
\centering
{\setlength{\tabcolsep}{2pt}
\begin{tabular}{|c|c|c|c|c|}
\hline
\bfseries Dataset & \bfseries ESPIRiT& \bfseries ESPIRiT &\bfseries  Nullspace & \bfseries Nullspace \\
\bfseries  & \bfseries & \bfseries + PISCO &\bfseries  & \bfseries + PISCO 
\\ 
\hline
Brain MPRAGE &1.8&0.1&47.2$^*$&0.1\\\hline
Brain TSE &3.2&0.1&127.0$^*$&0.1\\\hline
Knee TSE &3.5&0.1&35.2$^*$&0.1\\\hline
\end{tabular}
}
\label{tab:mem}
\end{table}

\section{Proposed Computational Improvements}\label{sec:5}

PISCO is comprised  of five separate computational approaches, which are independently described and individually evaluated in the first five subsections of this section (one subsection per approach).   Subsequently, we discuss the combined use of all five approaches in Sec.~\ref{sec:combo}.  

The following subsections (as well as the results already reported in the previous section) use common infrastructure to quantify the complexity of different computational approaches.  All methods were implemented in-house using MATLAB, and all computations were performed  on a desktop computer with an Intel Xeon E5-1603 2.8GHz quad-core CPU and 32GB RAM.  Computation times were determined by calculating the elapsed time (in seconds) between the start and end of the computation (i.e., the wall-clock time).  To minimize the potential for confounding effects, these computations were performed during quiescent periods (without other major software applications running in the foreground, and away from times when major background processes were scheduled to run).  To reduce the influence of random background activity, we repeated each computation at least five times and report the median.  We also evaluated the peak memory required for each implementation.  To illustrate performance in different scenarios, we performed computations for the Knee TSE dataset (which had the smallest number of channels with $Q=15$) using a smaller 5$\times$5 FIR filter support and the Brain TSE dataset (which had a larger number of channels with $Q=32$) using a larger 7$\times$7 FIR filter support.

In our empirical evaluations, we compare our acceleration techniques against what we call the ``Nullspace Baseline'' (NB) approach.  The NB corresponds to the naive nullspace-based algorithm from Sec.~\ref{sec:alg_nullspace} combined with the memory-saving approach that will be described in Sec.~\ref{sec:G_fast}.\footnote{As described in the previous section, we do not directly implement the naive nullspace method as the most straightforward implementation would exceed MATLAB's limits on the maximum array size.}

The different components of PISCO are described  below.

\subsection{FFT-Based Calculations to Determine the Nullspace of $\mathbf{C}$}\label{sec:fft_C}
 
 Many subspace-based methods require constructing the $\mathbf{C}$ matrix and then calculating $\mathbf{C}^H\mathbf{C}$ and its approximate nullspace (cf. steps 1 and 2 of the naive nullspace-based algorithm from Sec.~\ref{sec:alg_nullspace}). This calculation can sometimes be expensive because the $\mathbf{C}$ matrix can be quite large in scenarios where $P \gg Q|\Lambda|$.  Generally speaking, the value of $P$ will be larger whenever the FIR filter support $\Lambda$ is smaller and/or whenever the size of the calibration region is larger, and the acceleration approach described in this subsection will be particularly beneficial in those circumstances.
 
 Rather than working directly with $\mathbf{C}$ (and similar to previous work on structured low-rank MR image reconstruction  \cite{ongie2017fast, zhao2021high}), our approach is based on directly calculating the matrix $\mathbf{C}^H\mathbf{C}$ without ever calculating $\mathbf{C}$.  This can be beneficial because $\mathbf{C}$ and $\mathbf{C}^H\mathbf{C}$ share the same nullspace structure, while $\mathbf{C}^H\mathbf{C}$ will have fewer entries than $\mathbf{C}$ whenever $P > Q|\Lambda|$.   
 
 An efficient FFT-based approach for directly calculating $\mathbf{C}^H\mathbf{C}$ was originally introduced in Ref.~\cite{ongie2017fast} for the single-channel case, which was  later extendeded to the multichannel case in Ref.~\cite{zhao2021high}.  Our proposed approach follows the same basic principles though includes some improvements over the approach of Ref.~\cite{zhao2021high}, as described below. 
 
Based on the definition of $\mathbf{C}$ from Eq.~\eqref{eq:C}, the matrix $\mathbf{C}^H\mathbf{C}$ has $Q \times Q$ block structure, where the $(p,q)$th  block is given by $\mathbf{C}_p^H \mathbf{C}_q$ for $p,q \in \{1,\ldots,Q\}$. This is
  \begin{equation}
\mathbf{C}^H\mathbf{C}=\begin{bmatrix} 
\mathbf{C}_1^H \mathbf{C}_1 &  \!\cdots\! & \mathbf{C}_1^H \mathbf{C}_q &\!\cdots\! &\mathbf{C}_1^H\mathbf{C}_Q \\
\vdots &  \ddots & \vdots & \ddots &\vdots \\
\mathbf{C}_p^H \mathbf{C}_1 &  \!\cdots\! & \mathbf{C}_p^H \mathbf{C}_q &\!\cdots\! &\mathbf{C}_p^H\mathbf{C}_Q \\
\vdots &  \ddots & \vdots & \ddots &\vdots \\
\mathbf{C}_Q^H \mathbf{C}_1 &  \!\cdots\! & \mathbf{C}_Q^H \mathbf{C}_q &\!\cdots\! &\mathbf{C}_Q^H\mathbf{C}_Q \\
\end{bmatrix}.
  \end{equation}
  Construction of $\mathbf{C}^H\mathbf{C}$ can thus be performed efficiently if we have efficient methods for computing $\mathbf{C}_p^H \mathbf{C}_q$.  
  
  Note that efficient methods for computing $\mathbf{C}_p^H\mathbf{C}_q$ can potentially be obtained if we have efficient methods for computing matrix-vector multiplications of the form $\mathbf{C}_p^H\mathbf{C}_q \mathbf{v}$ for arbitrary vectors $\mathbf{v}\in \mathbb{C}^{|\Lambda|}$.   Specifically, the $i$th column of the matrix $\mathbf{C}_p^H\mathbf{C}_q$ can be computed using the matrix-vector multiplication $\mathbf{C}_p^H\mathbf{C}_q \mathbf{e}_i$, where $\mathbf{e}_i$ is the $i$th column of the identity matrix $\mathbf{I}_{|\Lambda|}$. This allows the entire $\mathbf{C}^H\mathbf{C}$ matrix to be calculated by calculating $\mathbf{C}_p^H\mathbf{C}_q \mathbf{e}_i$ for each $(p,q,i)$ combination with $p, q \in  \{1,\ldots,Q\}$,  and $i\in\{1,\ldots, |\Lambda|\}$.  
  
The convolution structure of the $\mathbf{C}_q$ matrices implies that, for any vector $\mathbf{v} \in \mathbb{C}^{|\Lambda|}$, the matrix-vector multiplication $\mathbf{C}_q\mathbf{v}$ can be expressed as first performing the simple convolution
\begin{equation}
 b[\mathbf{n}] = v[\mathbf{n}] \circledast s_q[\mathbf{n}] \triangleq \sum_{\mathbf{m} \in \Lambda} v[\mathbf{m}] s_q[\mathbf{n}-\mathbf{m}]
\end{equation}
(where $v[\mathbf{n}]$ is a simple reformatting of $\mathbf{v}$ and $\circledast$ denotes convolution) and then extracting the $P$ samples of $b[\mathbf{n}]$ corresponding to the $P$ values of $\mathbf{n}$ that were used in the construction of $\mathbf{C}$  (which can be viewed as masking  the convolution output).  Similarly, for any vector $\mathbf{u} \in \mathbb{C}^P$, the matrix-vector multiplication $\mathbf{C}_p^H \mathbf{u}$ can be expressed as reformatting the vector $\mathbf{u}$ into a sequence $u[\mathbf{n}]$, computing $u[\mathbf{n}] \circledast s_p^*[-\mathbf{n}]$, and finally extracting relevant samples from the convolution output, where $^*$ denotes complex conjugation.  

As described by Ongie and Jacob \cite{ongie2017fast}, the separate convolution operations that can be used to calculate $\mathbf{C}_q\mathbf{v}$ and $\mathbf{C}^H_p\mathbf{u}$ can each be efficiently computed (without approximation) using standard FFT-based methods.  Moreover, if the masking operations are neglected (which typically only has a minor effect on the final results \cite{ongie2017fast}), then the composite matrix-vector multiplication  $\mathbf{C}_p^H\mathbf{C}_q \mathbf{v}$ can be accurately approximated by calculating $v[\mathbf{n}] \circledast (s_q[\mathbf{n}] \circledast s_p^*[-\mathbf{n}])$ and extracting relevant samples from the convolution output.  If zero-padded FFTs of $s_q[\mathbf{n}]$ and $s_p[\mathbf{n}]$ have been precomputed, then this convolution can be easily computed by taking the zero-padded FFT of $v[\mathbf{n}]$, multiplying the result with the precomputed FFTs of $s_q[\mathbf{n}]$ and $s_p^*[-\mathbf{n}]$,\footnote{Note that the FFT of $s_p^*[-\mathbf{n}]$ can be easily retrieved from the FFT of $s_p[\mathbf{n}]$ using the conjugation property of the Fourier transform.} and then taking an inverse FFT.  

The steps described above closely resemble the multichannel approach of Zhao \emph{et al.} \cite{zhao2021high}.  In our implementation (and similar to the original single-channel work of Ongie and Jacob \cite{ongie2017fast}), we additionally use the fact that if $\mathbf{v} = \mathbf{e}_i$, then the corresponding sequence representation $v[\mathbf{n}]$  will be equal to $\delta[\mathbf{n} - \mathbf{n}_i]$, where $\delta[\cdot]$ is the Kronecker delta function and $\mathbf{n}_i \in \mathbb{Z}^D$ is the $i$th element of $\Lambda$.   As a result, the matrix-vector multiplications $\mathbf{C}_p^H\mathbf{C}_q \mathbf{e}_i$ for different choices of $i$ can be simply approximated as convolutions of the form $\delta[\mathbf{n}-\mathbf{n}_i] \circledast (s_q[\mathbf{n}] \circledast s_p^*[-\mathbf{n}])$ with different choices of $\mathbf{n}_i$.  Importantly, the sifting property of the Kronecker delta allows us to retrieve the results of $\delta[\mathbf{n}-\mathbf{n}_i] \circledast (s_q[\mathbf{n}] \circledast s_p^*[-\mathbf{n}])$ for every choice of $\mathbf{n}_i$ by simply shifting the output of $\delta[\mathbf{n}] \circledast (s_q[\mathbf{n}] \circledast s_p^*[-\mathbf{n}]) = s_q[\mathbf{n}] \circledast s_p^*[-\mathbf{n}]$ by a number of samples according to $\mathbf{n}_i$.  This means that it is not necessary to separately calculate distinct convolution results for each $i\in \{1,\ldots,|\Lambda|\}$. Instead, it suffices to calculate a single convolution $s_q[\mathbf{n}] \circledast s_p^*[-\mathbf{n}]$ per $(p,q)$ pair and then construct approximations of the columns of $\mathbf{C}_p^H\mathbf{C}_q$ by simple shifts of the convolution output!  

A close approximation of $\mathbf{C}^H\mathbf{C}$ can thus be obtained by precomputing zero-padded FFTs of each $s_q[\mathbf{n}]$ for $q=1,\ldots,Q$ (a total of $Q$ FFTs) and then multiplying the results together and computing inverse FFTs for each $(p,q)$ pair (a total of $Q^2$ additional FFTs).  Once this approximation of $\mathbf{C}^H\mathbf{C}$ is obtained, it is straightforward to calculate its approximate nullspace using standard SVD algorithms.  

The computation associated with this acceleration approach scales with $Q^2$, so is expected to be more advantageous in situations where the number of channels is small, and may not be beneficial in situations where $Q$ is large. And as already mentioned, this approach is expected to be more beneficial in situations where $P > Q|\Lambda|$, and is less likely to be beneficial in situations where $P < Q |\Lambda|$.   

\begin{figure}[t]
 \centering 
 \begin{minipage}{0.3\linewidth}
 \renewcommand\thesubfigure{(a)}
\subfigure[]{\includegraphics[width=1.2\linewidth]{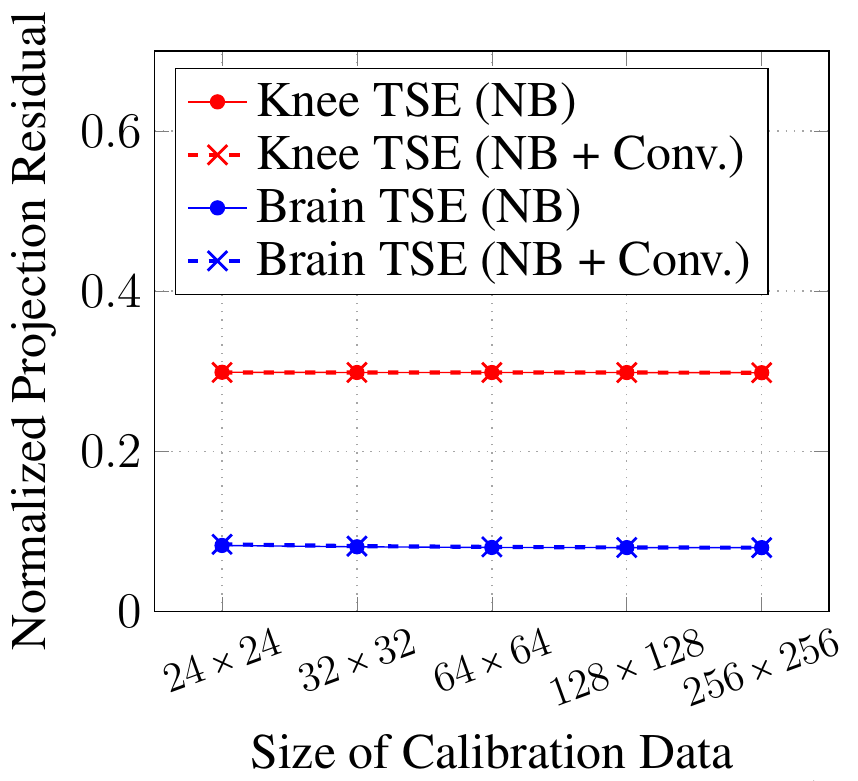}} 
 \end{minipage}
 \hspace{5mm}
 \begin{minipage}{0.3\linewidth}
 \vspace{2.5mm}
 \renewcommand\thesubfigure{(b)}
\subfigure[]{\includegraphics[width=1.15\linewidth]{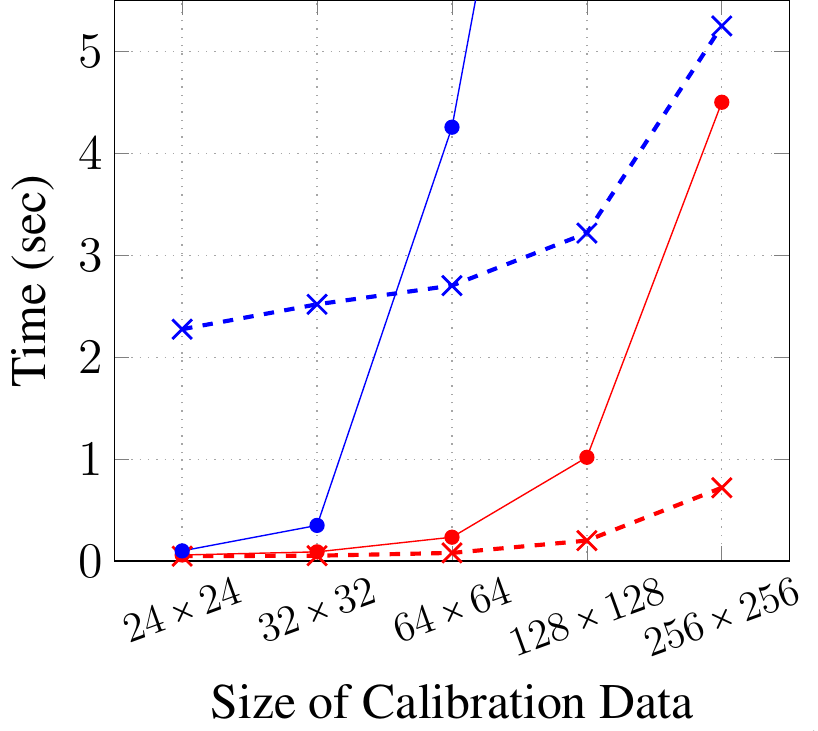}} 
 \end{minipage}
 \caption{  Evaluation of the convolutional acceleration from Sec.~\ref{sec:fft_C}. }
 \label{fig:plots_T2_SCO}
 \end{figure}

Our empirical results for this acceleration approach are shown in Fig. \ref{fig:plots_T2_SCO}, and are consistent with our theoretical expectations.  While our FFT-based approach does involve a small amount of approximation, the effects of this approximation are observed to be relatively minor, and the normalized projection residual is not substantially changed.  On the other hand, we see substantial (though somewhat nuanced) changes in computation time.  For large calibration regions (corresponding to situations where $P\gg Q|\Lambda|$), we consistently observe that the convolutional acceleration outperforms the Nullspace Baseline.  However, when the calibration region is small, the convolutional acceleration is only better than the Nullspace Baseline for the Knee TSE data, but not for the Brain TSE data.  This is consistent with expectations, as the Knee TSE data has both larger $P$ than the Brain TSE data (due to the use of a smaller FIR filter support $\Lambda$) and smaller $Q$.  While there are subtle differences in memory usage, these differences were not substantial  and are not shown.
  
 \subsection{Ellipsoidal FIR Filter Support }\label{sec:ellip}
 When constructing the $\mathbf{C}$ matrix, methods like ESPIRiT have traditionally chosen the FIR filter support $\Lambda$ to have a rectangular shape (e.g., in the 2D case, a $(2\tau+1)$\,\!$\times$\,\!$(2\tau+1)$ rectangular filter support is obtained when using $\Lambda = \{\mathbf{n} \in \mathbb{Z}^2: \|\mathbf{n}\|_\infty \leq \tau\}$).  However, previous work has shown that the corners of the rectangular filter have a negligible impact for methods that rely on the $\mathbf{C}$ matrix \cite{lobos2022shape}, and that removing these unimportant corners (i.e., simplifying the rectangle into an ellipsoid using $\Lambda = \{\mathbf{n} \in \mathbb{Z}^2: \|\mathbf{n}\|_2 \leq \tau\}$) offers substantial computational advantages with negligible impact on the final results.  This same approach is adopted in PISCO.
 
\begin{figure}[t]
 \centering 
 \begin{minipage}{0.3\linewidth}
 \renewcommand\thesubfigure{(a)}
\subfigure[]{\includegraphics[width=1.2\linewidth]{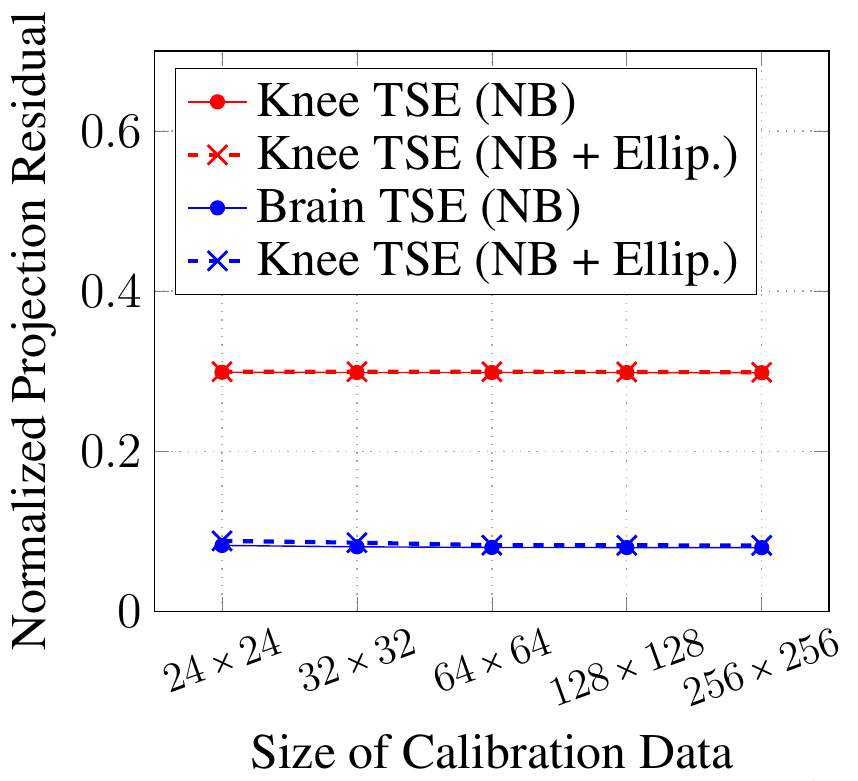}} 
 \end{minipage}
  \hspace{5mm}
  \begin{minipage}{0.3\linewidth}
 \vspace{2.3mm}
 \renewcommand\thesubfigure{(b)}
\subfigure[]{\includegraphics[width=1.15\linewidth]{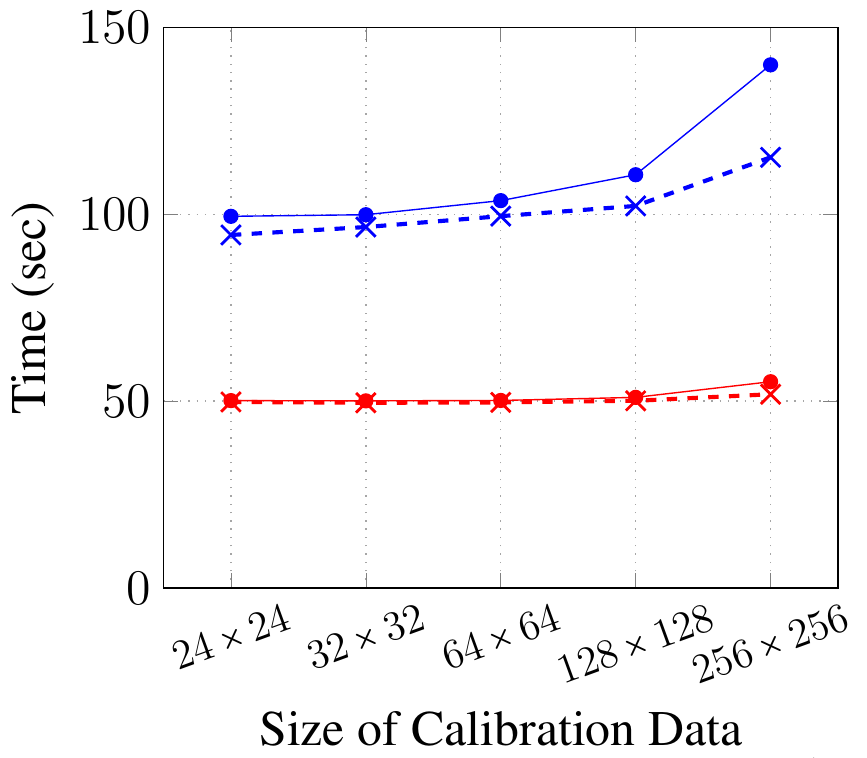}} 
 \end{minipage}
 \caption{Evaluation of the ellipsoidal acceleration from Sec.~\ref{sec:ellip}.}
 \label{fig:plots_T2_S}
 \end{figure}

 Our empirical results for this acceleration approach are shown in Fig.~\ref{fig:plots_T2_S}.  Although this acceleration approach does slightly change the final sensitivity maps compared to the Nullspace Baseline, the differences are not  substantial as can be observed from the normalized projection residual.  However, the acceleration approach does provide a slight but noticeable benefit to the total computation time when calculating sensitivity maps, particularly as the size of the calibration region grows.  Although this acceleration approach also results in memory benefits as previously described \cite{lobos2022shape}, these benefits were not very noticeable (and are not shown) since the steps involving the $\mathbf{C}$ matrix are not the most memory-consuming steps of the Nullspace Baseline.
 
\subsection{Replacing $\mathbf{H}(\mathbf{x})$ with $\mathbf{G}(\mathbf{x})$}\label{sec:G_fast}
As already noted, constructing  $\mathbf{H}(\mathbf{x})$  (i.e., constructing one $R \times Q$ matrix for every one of the $N$ spatial locations in the image) can require a substantial amount of memory, and can impede implementation of the naive nullspace-based algorithm.  While there are simple ways to bypass this problem (i.e., only computing $\mathbf{H}(\mathbf{x})$ for one voxel at a time instead of storing the matrix values for all voxels simultaneously), this can result in a loop over all voxels that can be very slow in the common case where the number of voxels is large.

Our proposed approach is based on replacing $\mathbf{H}(\mathbf{x})$ with 
\begin{equation}
 \mathbf{G}(\mathbf{x})  \in \mathbb{C}^{Q \times Q} \triangleq (\mathbf{H}(\mathbf{x}))^H \mathbf{H}(\mathbf{x}).
\end{equation}
Similar to the arguments we made for preferring $\mathbf{C}^H\mathbf{C}$ over  $\mathbf{C}$ in Sec.~\ref{sec:C}, the matrix $\mathbf{G}(\mathbf{x}) \in \mathbb{C}^{Q \times Q}$ shares the same nullspace structure with $\mathbf{H}(\mathbf{x})\in \mathbb{C}^{R \times Q}$, while $\mathbf{G}(\mathbf{x})$ will be substantially smaller when the number of nullspace filters $R$ is much larger than the number of channels $Q$ (which is a common occurence as can be seen from the large nullspaces observed in Fig.~\ref{fig:plots_C_sv}).  Directly calculating  $\mathbf{G}(\mathbf{x})$ instead of $\mathbf{H}(\mathbf{x})$ is thus expected to reduce memory utilization by at least a factor of $R/Q$, which is often substantial (e.g., the value of $R/Q$ ranges from between 20 to 50 for the datasets and parameter settings considered in this paper!).

A further observation that underlies our proposed approach is that $\mathbf{G}(\mathbf{x})$ is structured, with entries that can be written as

\begin{equation}
\begin{split}
 [\mathbf{G}(\mathbf{x})]_{pq} &= \sum_{r=1}^R h_r^*(\mathbf{x},p) h_r(\mathbf{x},q)\\
 &\hspace{-0.4in}=  \sum_{s = 1}^{|\Lambda|} \sum_{t =1}^{|\Lambda|}  e^{i2\pi (\mathbf{n}_s - \mathbf{m}_t)^T\mathbf{x}} \left(\sum_{r=1}^R h_r^*[\mathbf{m}_t,p]h_r[\mathbf{n}_s,q]\right). \label{eq:24}
 \end{split}
\end{equation}
A naive calculation of $\mathbf{G}(\mathbf{x})$ at every spatial location using this formula would use $Q^2|\Lambda|^2(R+1)N$ multiplications, which could be quite expensive.

To proceed further, let  $\mathbf{W} \in \mathbb{C}^{Q|\Lambda| \times Q|\Lambda|}$ be defined as $\mathbf{W} = \sum_{r=1}^R \mathbf{h}_r\mathbf{h}_r^H$,  
where $\mathbf{h}_r$ is the $r$th approximate nullspace vector of $\mathbf{C}$ as previously defined in Sec.~\ref{sec:C}. Note that $\mathbf{W}$ is easily calculated using $Q^2|\Lambda|^2 R$ multiplications.  It is straightforward to observe that $\mathbf{W}$ has a $Q \times Q$ block structure, where we denote the $(q,p)$th block by $\mathbf{W}_{qp} \in \mathbb{C}^{|\Lambda|\times\Lambda|}$ for $p,q \in \{1,\ldots,Q\}$. Specifically, 
\begin{equation}
 \mathbf{W}  = \begin{bmatrix} \mathbf{W}_{11} & \cdots & \mathbf{W}_{1p} & \cdots & \mathbf{W}_{1Q} \\ 
 \vdots & \ddots & \vdots & \ddots & \vdots \\
 \mathbf{W}_{q1} & \cdots & \mathbf{W}_{qp} & \cdots & \mathbf{W}_{qQ} \\
 \vdots & \ddots & \vdots & \ddots & \vdots \\
 \mathbf{W}_{Q1} & \cdots & \mathbf{W}_{Qp} & \cdots & \mathbf{W}_{QQ}
 \end{bmatrix}.
\end{equation}
The entry in the $s$th row and $t$th column of  $\mathbf{W}_{qp}$ is given by $[\mathbf{W}_{qp}]_{st} = \sum_{r=1}^R h_r^*[\mathbf{m}_t,p]h_r[\mathbf{n}_s,q]$, which corresponds to one of the terms appearing in Eq.~\eqref{eq:24}.  Using $w_{qp}^s[\mathbf{m}_t]$  to denote the $s$th row of $\mathbf{W}_{qp}$ reformatted as a sequence and leveraging the shift property of the discrete Fourier transform, Eq.~\eqref{eq:24} simplifies to
\begin{equation}
 [\mathbf{G}(\mathbf{x})]_{pq}  = \sum_{\mathbf{m} \in \mathbb{Z}^D} \left( \sum_{s = 1}^{|\Lambda|}   w_{qp}^s[\mathbf{m}+\mathbf{n}_s]\right)  e^{-i2\pi \mathbf{m}^T\mathbf{x}}.
\end{equation}
Importantly, the inner summation over $s$ does not require any additional multiplications beyond the $Q^2|\Lambda|^2R$ multiplications that were already used in the formation of $\mathbf{W}$, while the outer summation can be easily evaluated for every spatial location using a zero-padded FFT ($O(N\log_2 N)$ multiplications). This calculation needs to be repeated for every $(p,q)$ combination,  such that the total number of complex multiplications required for our proposed approach is thus $O(Q^2 (N\log_2 N +|\Lambda|^2R))$. This can be substantially smaller than the $Q^2|\Lambda|^2(R+1)N$ multiplications needed for the naive calculation of $\mathbf{G}(\mathbf{x})$. 

\begin{figure}[t]
 \centering 
 \begin{minipage}{0.55\linewidth} 
\subfigure{\includegraphics[width=1.15\linewidth]{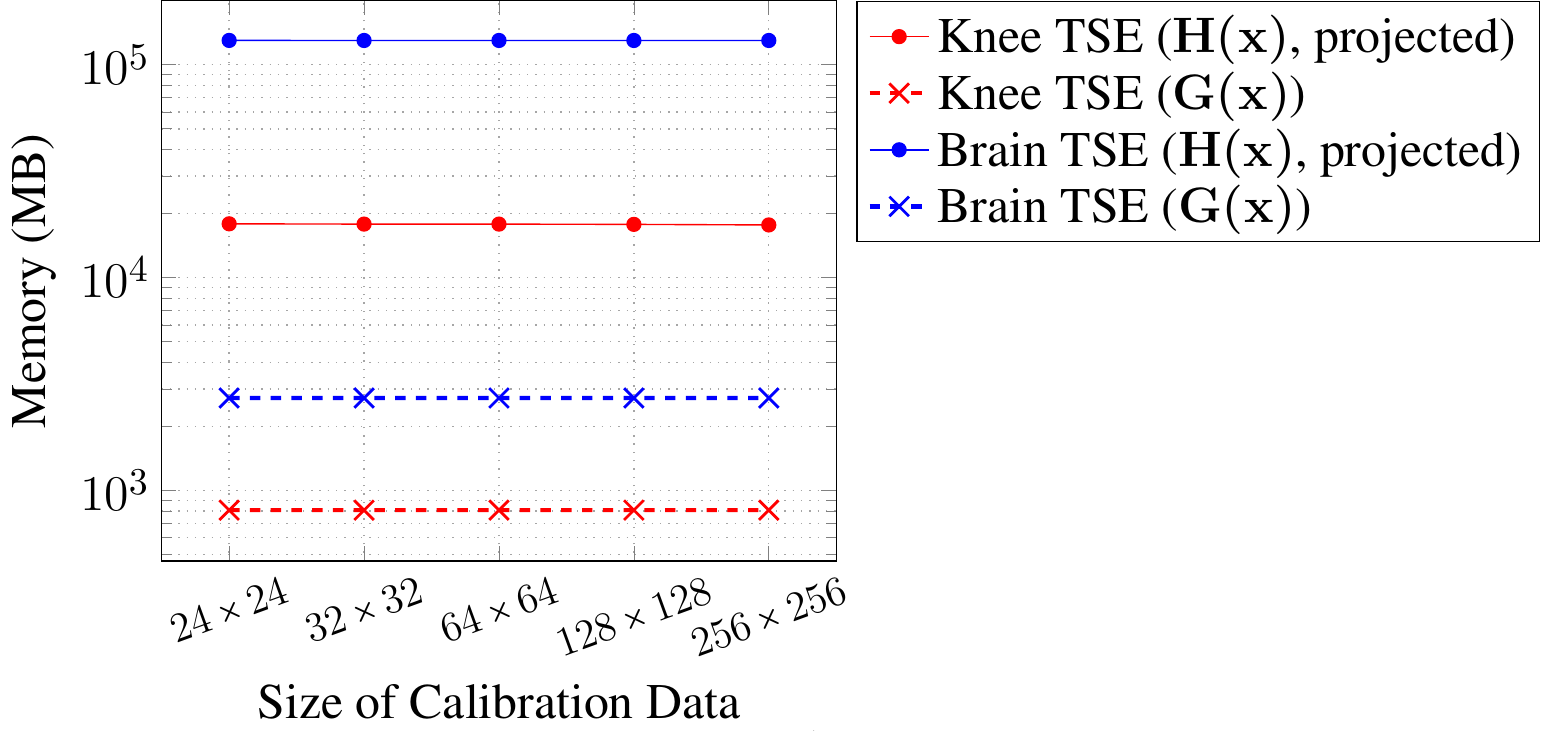}} 
 \end{minipage}
 \caption{Evaluation of the $\mathbf{G}(\mathbf{x})$ calculation from Sec.~\ref{sec:G_fast}.}
 \label{fig:G}
 \end{figure}

Since we could not implement the naive nullspace-based algorithm using $\mathbf{H}(\mathbf{x})$, our new ability to calculate $\mathbf{G}(\mathbf{x})$ is an enabling technology for the nullspace-based algorithm.  Figure~\ref{fig:G} shows a comparison between the memory utilized by our proposed $\mathbf{G}(\mathbf{x})$ approach compared against our projections of the memory required for $\mathbf{H}(\mathbf{x})$.  As expected, there are  orders-of-magnitude improvements in required memory.

 \subsection{Smoothness-Based Interpolation}\label{sec:interp}

The fact that subspace-based methods require manipulating matrices like $\mathbf{H}(\mathbf{x})$ or $\mathbf{G}(\mathbf{x})$ at every spatial location $\mathbf{x}$ can also be burdensome if the number of spatial locations $N$ is large.  
 Using the well-known smoothness characteristics of sensitivity maps, we observe that instead of estimating sensitivity maps directly for the nominal high-resolution voxel grid, it should be sufficient to compute $\mathbf{G}(\mathbf{x})$ and estimate sensitivity maps on a low-resolution voxel grid, and then interpolate back to the desired spatial resolution. Our implementation relies on periodic sinc interpolation, which can be implemented efficiently using zero-padded FFTs.  For the implementation reported in this work, we used a low-resolution voxel grid that was heuristically chosen to be 24 samples bigger than the calibration region along each dimension (i.e., for a 24$\times$24 calibration region, we used a 48$\times$48 low-resolution voxel grid).   This choice worked well for the datasets considered in this paper, although alternative heuristics may lead to even better results, and the optimal selection of interpolation factors is likely to be application-dependent.
 
 For interpolation to be successful, it is necessary  for the low-resolution sensitivity map estimates to be smooth, which is potentially confounded by the ambiguities described in Sec.~\ref{sec:ambiguity}, including the inherent phase-ambiguities associated with the SVD of $\mathbf{G}(\mathbf{x})$ which can  result in  quasi-random non-smooth phase characteristics.  To avoid this issue, we choose a scaling function $\alpha(\mathbf{x})$ to provide smooth phase characteristics. Specifically, we use the initial sensitivity map estimates (which often have arbitrary nonsmooth phase) to perform  coil-combination of low-resolution (Gaussian-apodized) Fourier reconstructions of the calibration data, and then choose $\alpha(\mathbf{x})$ to cancel the phase of the coil-combined reconstruction (so that the coil-combined reconstruction becomes real-valued).  This approach is based on the assumption that the underlying image $\rho(\mathbf{x})$ does not have substantial phase variations such that the observed phase can be largely be ascribed to the sensitivity maps, which are presumed smooth.  This approach worked well for the datasets considered in this paper, although more care may be needed in cases where the image phase is expected to be more complicated.
 
\begin{figure}[t]
 \centering 
 \hspace{-4mm}
 \begin{minipage}{0.3\linewidth}
 \renewcommand\thesubfigure{(a) }
\subfigure[]{\includegraphics[width=1.15\linewidth]{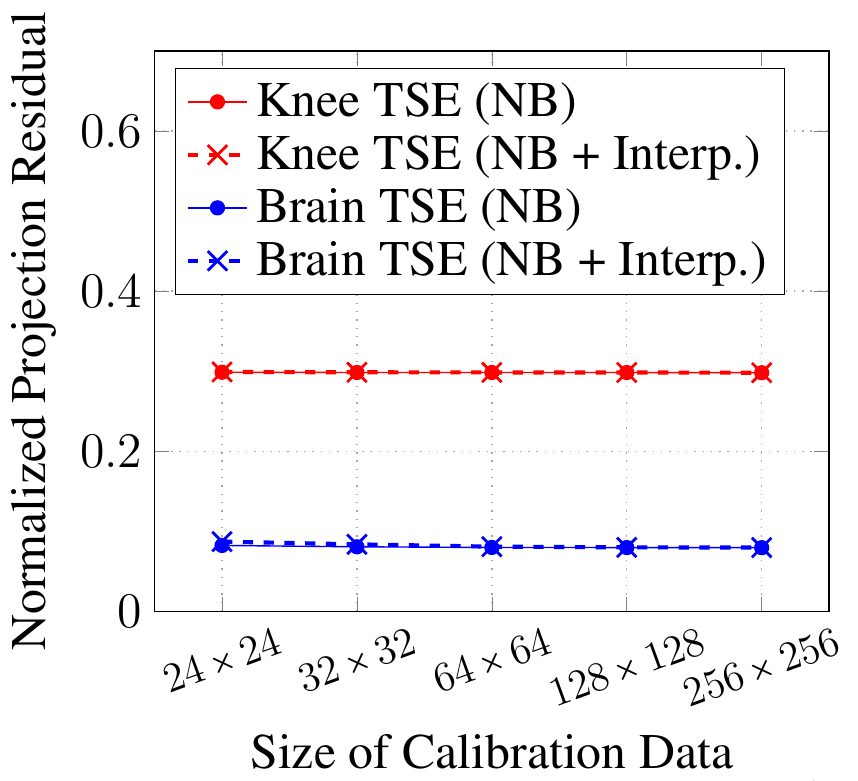}}
 \end{minipage}
 \hspace{1mm}
 \begin{minipage}{0.3\linewidth}
  \vspace{1.8mm}
 \renewcommand\thesubfigure{(b) }
\subfigure[]{\includegraphics[width=1.15\linewidth]{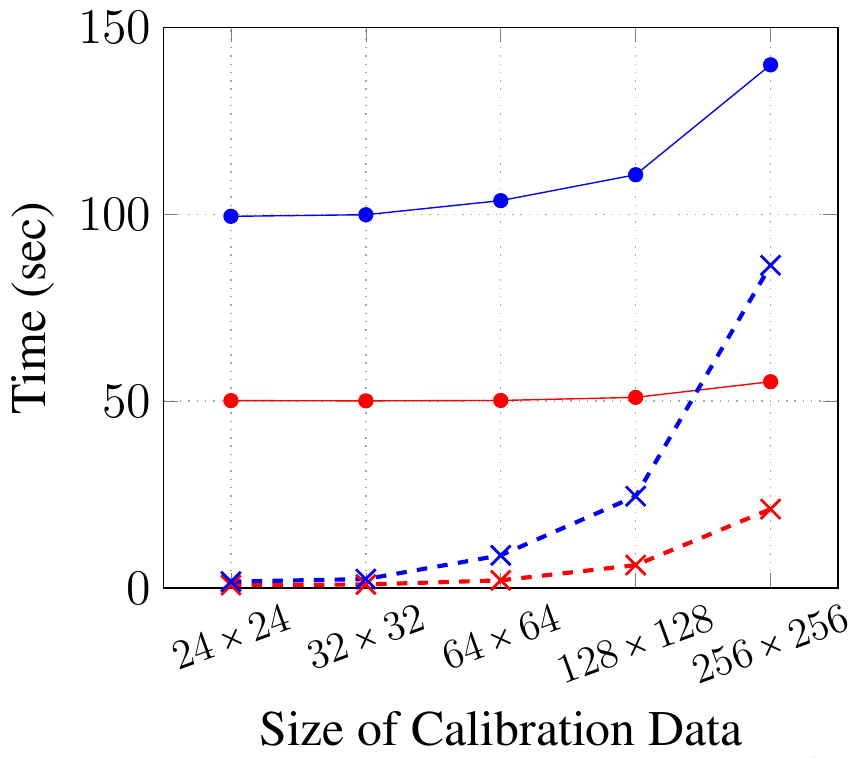}}
\end{minipage}
 \hspace{1mm}
 \begin{minipage}{0.3\linewidth}
   \vspace{1.8mm}
 \renewcommand\thesubfigure{(c) }
\subfigure[]{\includegraphics[width=1.15\linewidth]{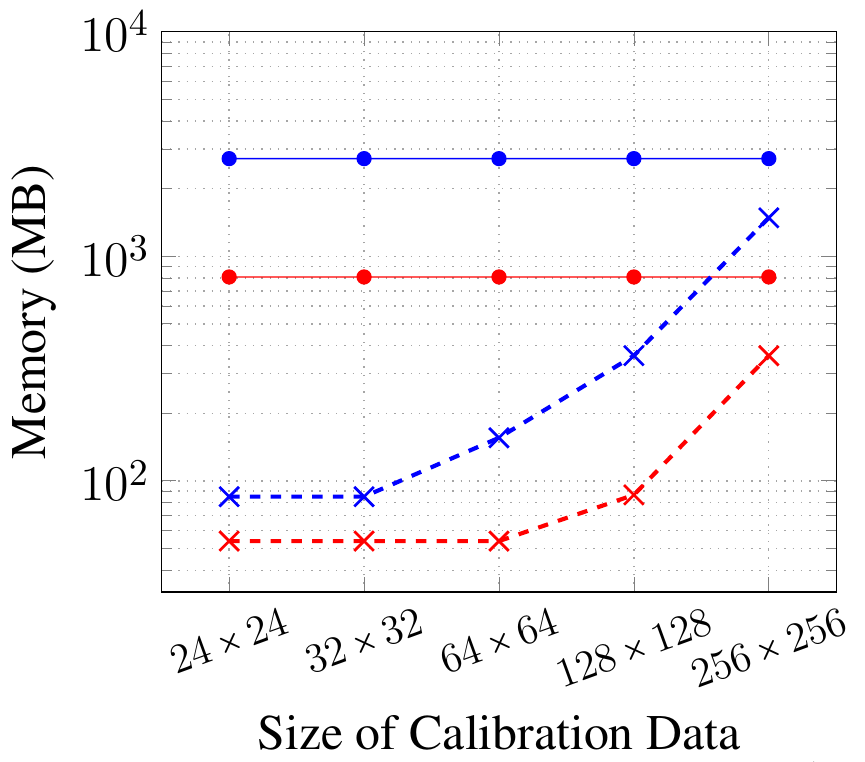}}
\end{minipage}
 \caption{Evaluation of the interpolation approach from Sec.~\ref{sec:interp}.   }
 \label{fig:plots_T2_I}
 \end{figure}

 Our empirical results for interpolation are reported in Fig.~\ref{fig:plots_T2_I}, which demonstrates that interpolation leads to substantial reductions in the total computation time and memory utilization with only minor effects on the normalized projection residual.

 \subsection{Power Iteration} \label{sec:PI}
 One of the most expensive steps of a subspace-based approach is the calculation of SVDs of $\mathbf{H}(\mathbf{x})$ or $\mathbf{G}(\mathbf{x})$ at every spatial location.  Rather than performing full SVDs, we propose to use iterative methods to estimate only the right singular vector associated with the smallest singular value.  For example, this can be done using classical power iteration \cite{golub1996}, and it is also straightforward to modify the iterative procedure to perform this estimation process for all spatial locations simultaneously in parallel using vectorization techniques, thereby avoiding a loop over spatial locations.

Fast convergence of power iteration is expected when there is a big difference in magnitude between the smallest two singular values of $\mathbf{G}(\mathbf{x})$ \cite{golub1996}, which we have observed is usually the case (cf. Fig.~\ref{fig:sv_H}).   As a result, we only need to use a few iterations to achieve suitable convergence.  However, in cases where multiple sensitivity maps are required to describe a specific spatial location (e.g., when aliasing occurs within the FOV or in the presence of substantial partial voluming), then the smallest two singular values may have similar magnitudes and power iteration may be slower to converge. In such cases, the use of more advanced iterative algorithms (e.g., the Lanczos algorithm) may be preferable.

\begin{figure}[t]
 \centering 
 \begin{minipage}{0.3\linewidth}
 \renewcommand\thesubfigure{(a)}
\subfigure[]{\includegraphics[width=1.15\linewidth]{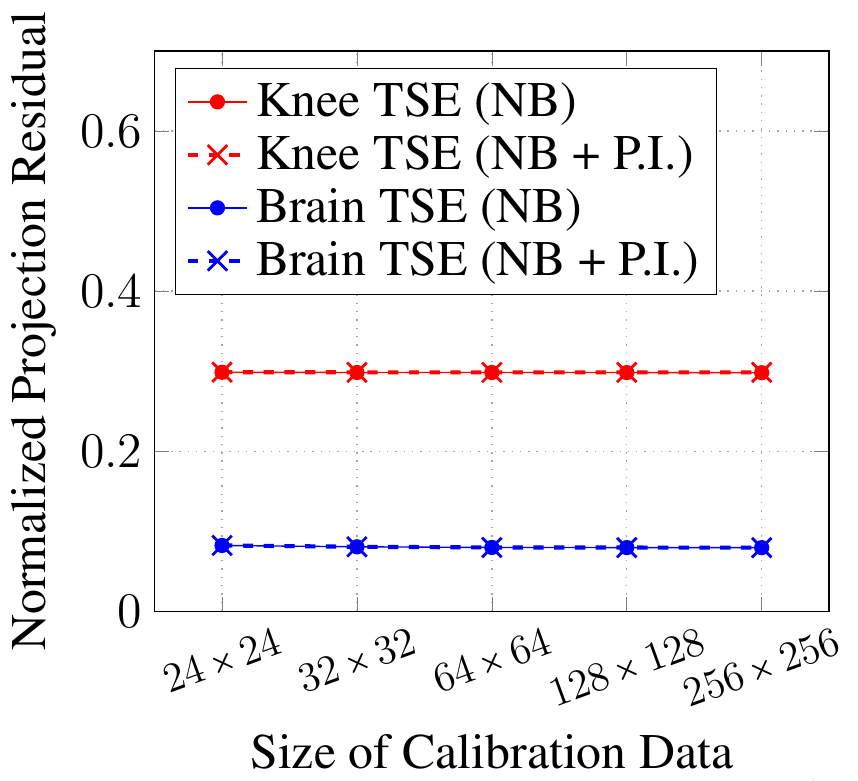}} 
 \end{minipage}
 \hspace{5mm}
 \begin{minipage}{0.3\linewidth}
 \vspace{1.7mm}
 \renewcommand\thesubfigure{(b)}
\subfigure[]{\includegraphics[width=1.15\linewidth]{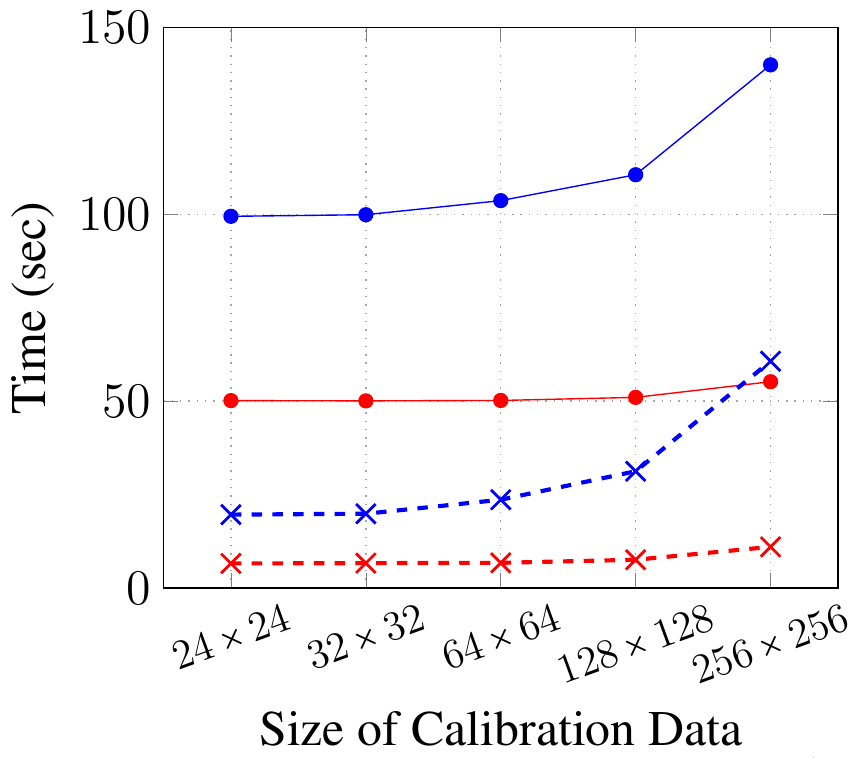}} 
 \end{minipage}
 \caption{ Evaluation of the power iteration approach from Sec.~\ref{sec:PI}.  }
 \label{fig:plots_T2_P}
 \end{figure}

Our empirical results for power iteration (which used a fixed number of 10 iterations) are shown in Fig.~\ref{fig:plots_T2_P}, and demonstrate substantial improvements in the total computation time relative to the Nullspace Baseline with only minor changes in the normalized projection residual. The use of power iteration did not have a major impact on memory utilization (not shown).

\begin{figure}[t]
 \centering 
 \hspace{-4mm}
 \begin{minipage}{0.3\linewidth}
 \renewcommand\thesubfigure{(a) }
\subfigure[]{\includegraphics[width=1.15\linewidth]{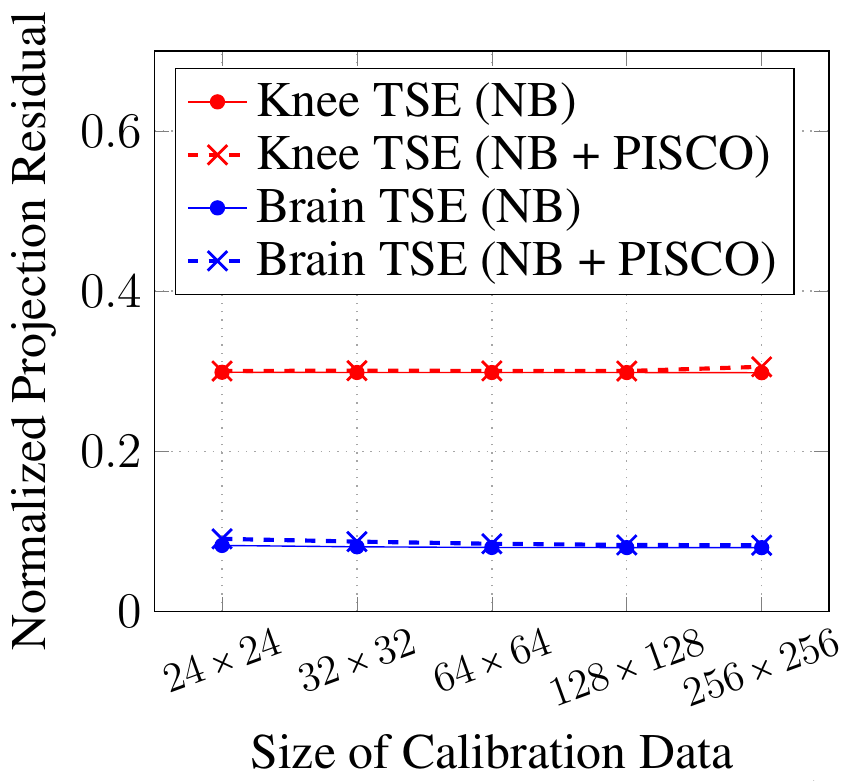}}
 \end{minipage}
 \hspace{1mm}
 \begin{minipage}{0.3\linewidth}
  \vspace{1.8mm}
 \renewcommand\thesubfigure{(b) }
\subfigure[]{\includegraphics[width=1.15\linewidth]{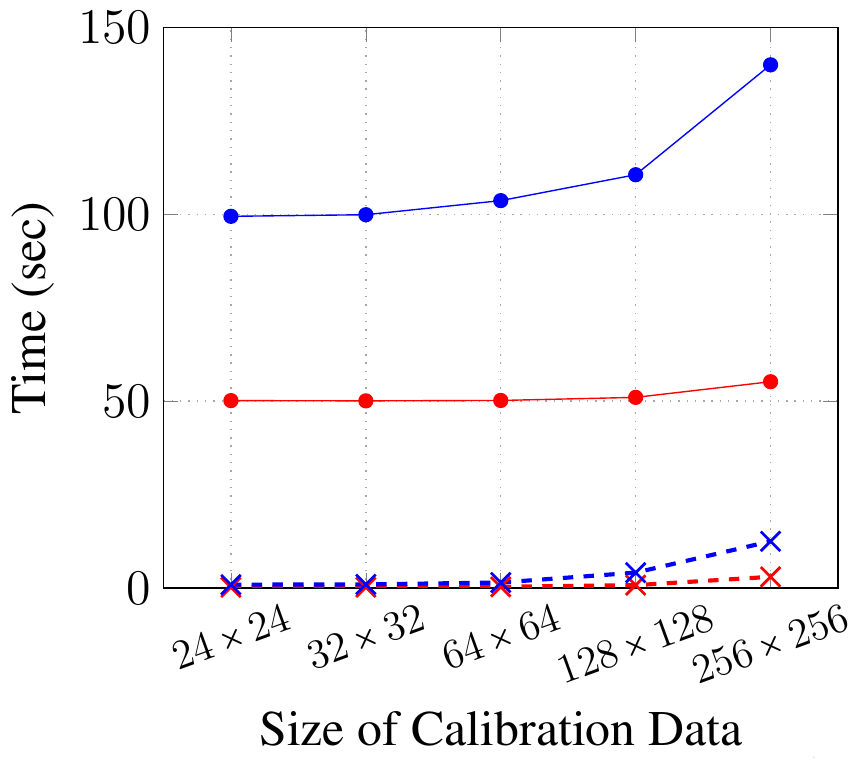}}
\end{minipage}
 \hspace{1mm}
 \begin{minipage}{0.3\linewidth}
   \vspace{1.8mm}
 \renewcommand\thesubfigure{(c) }
\subfigure[]{\includegraphics[width=1.15\linewidth]{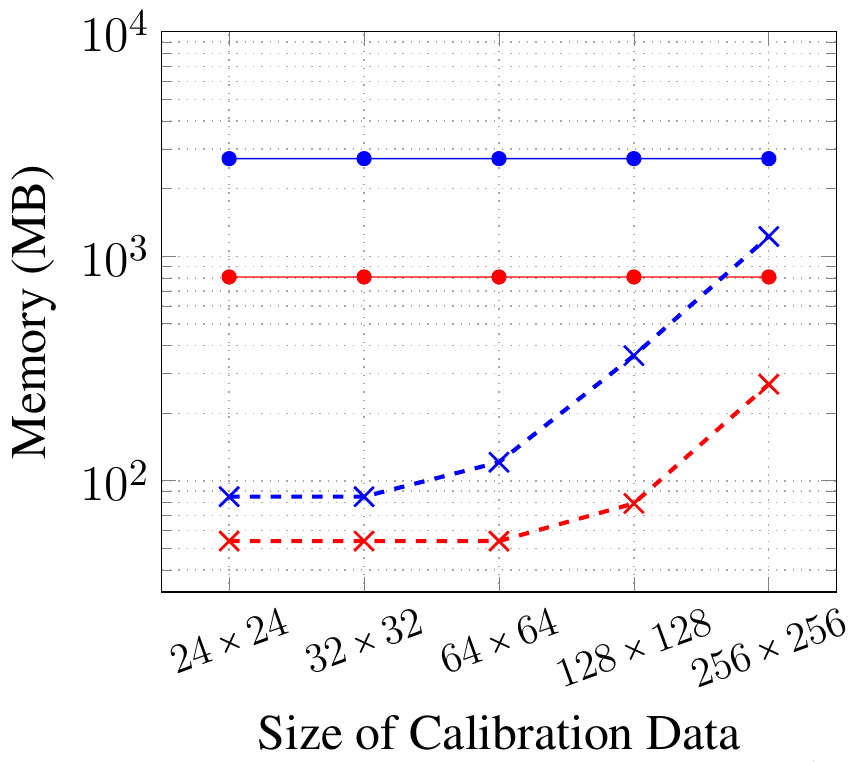}}
\end{minipage}
 \caption{ Comparison between the Nullspace Baseline and the nullspace-based algorithm with all five PISCO components.  }
 \label{fig:plots_T2_NAIVE_PISCO}
 \end{figure}

 \subsection{All Components of PISCO Combined}\label{sec:combo}
The previous subsections demonstrated that each of the individual components of PISCO can separately provide computational advantages.  In this section, we demonstrate that the combination of all five PISCO approaches leads to even further improvements.  Note that while some acceleration techniques did not provide improvements in every situation (e.g., the FFT-based calculation of the nullspace of $\mathbf{C}$ from Sec.~\ref{sec:C}), we still always used all five techniques together for simplicity.  Even better performance would have been obtained if we selectively turned off acceleration techniques in regimes where they were not useful.

\begin{figure}[t]
 \centering 
\includegraphics[width=0.4\textwidth]{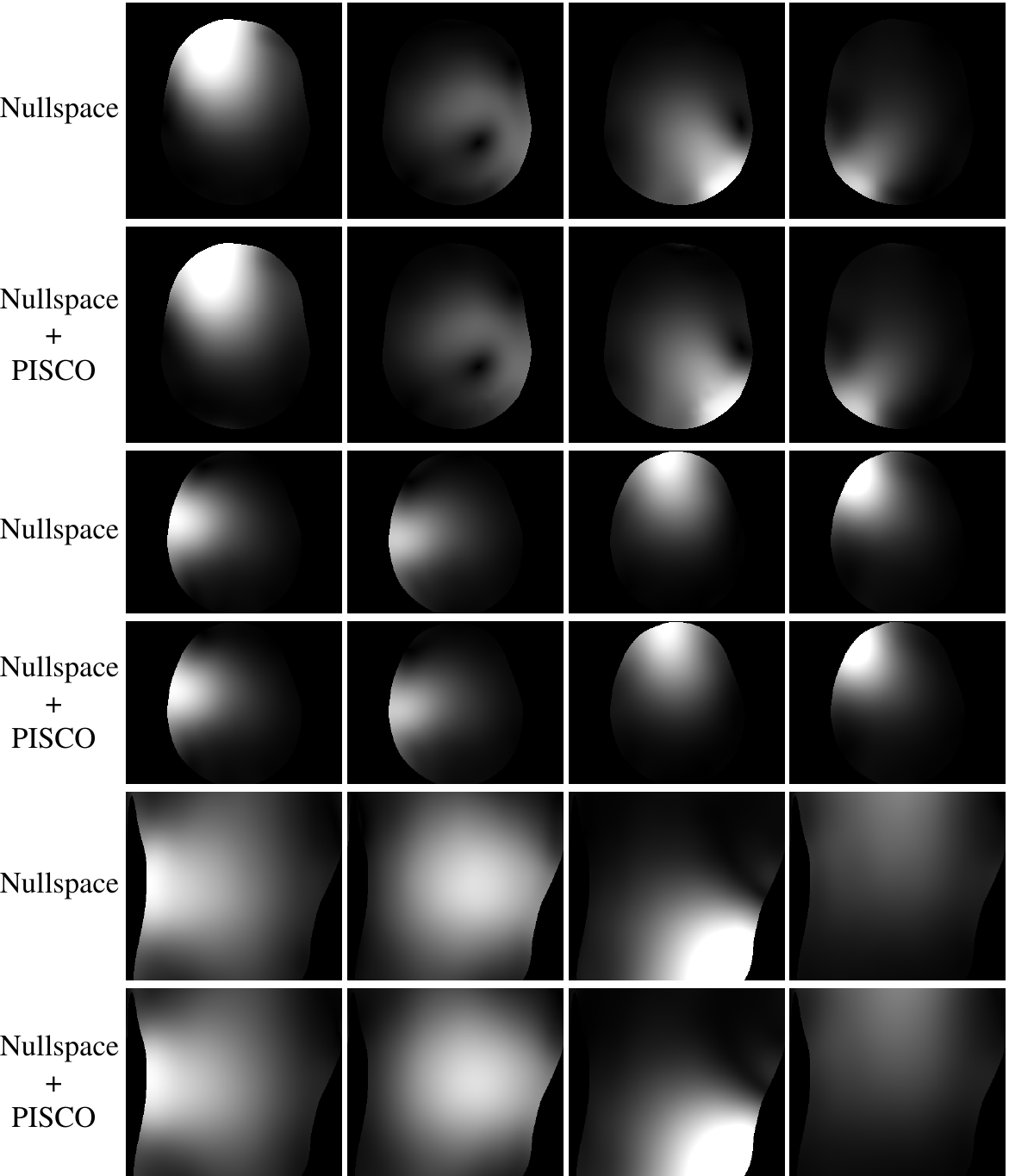}
 \caption{ Estimated sensitivity maps using the nullspace-based algorithm  with and without PISCO.  Each column shows the magnitude of the estimated sensitivity map for a different representative coil.  The first two rows correspond to the Brain MPRAGE dataset, the middle two rows correspond to the Brain TSE dataset, while the bottom two rows correspond to the Knee TSE dataset. }
 \label{fig:maps_comp}
 \end{figure}

 Figure~\ref{fig:plots_T2_NAIVE_PISCO} compares the results of the Nullspace Baseline against the nullspace-based method with PISCO.  In all scenarios, we observe consistently substantial improvements in computation time (up to $\sim$100$\times$ acceleration in some cases) and substantial improvements in memory usage (even though the Nullspace Baseline already makes use of our $\mathbf{G}(\mathbf{x})$ computation, which is one of the most memory-saving components of PISCO).   Figure~\ref{fig:maps_comp} shows representative sensitivity maps, which qualitatively demonstrates the strong similarity between the sensitivity maps obtained with both approaches.  Detailed numbers for 32$\times$32 calibration data were already presented in Tables~\ref{tab:nrmse}-\ref{tab:mem} and support the same conclusions.

 \section{Discussion and Conclusions}\label{sec:disc}
 
 This work presented a new theoretical framework for subspace-based sensitivity map estimation based on principles from the literature on linear predictability and structured low-rank modeling.  This led to a nullspace-based sensitivity map estimation method that is mathematically equivalent to ESPIRiT, but which has a distinct and complementary theoretical explanation that will be potentially more intuitive to some readers.  In addition, we introduced the set of PISCO computational accelerations for subspace-based sensitivity map estimation that enable substantial reductions in computation time (up to $\sim$100$\times$ in some of our examples) without a substantial loss in estimation quality.  The PISCO class of techniques is not only useful for our nullspace-based algorithm, but also offers substantial improvements for other subspace-based sensitivity map estimation methods like ESPIRiT.  While we have not demonstrated further extensions to applications beyond sensitivity map estimation, we expect that some of the PISCO techniques may also enable computational improvements for other types of similar computations, including subspace-based adaptive coil combination \cite{walsh2000adaptive}, structured low-rank image reconstruction \cite{zhang2011, shin2014, haldar2014low, Haldar_2015,haldar2015autocalibrated, jin2015a, ongie2016, haldar2019,jacob2020}, and other patch-based low-rank modeling methods that require the simultaneous SVDs of a large number of small-scale matrices \cite{trzasko2011a,veraart2016,cordero-grande2019}.
 
 One of the optional steps given in the nullspace-based algorithm in Sect. \ref{sec:alg_nullspace}, corresponds to choosing a scaling function $\alpha(\mathbf{x})$ with useful properties to normalize the estimated sensitivity maps. In the experiments shown in this work we have chosen this scaling function such that the magnitude of the estimated sensitivity maps is normalized following a sum-of-squares approach, and the phase is referenced with respect to the phase of one of the estimated sensitivity maps \cite{uecker2014}. However, this type of scaling function is not necessarily optimal, and different applications can be benefited from other normalization approaches.
 
 It should be noted that the results presented in this paper were all based on in-house MATLAB implementations. We do not expect MATLAB computation times to be competitive with computation times achieved using software written in programming languages like C, and expect that computational improvements will easily be achieved  by implementing PISCO in another programming language. Interestingly, we have anecdotally compared a C-based implementation of ESPIRiT (i.e., the implementation available from the BART software package \cite{uecker2015berkeley}) against our PISCO-based MATLAB implementation of ESPIRiT using the Brain TSE dataset with 24$\times$24 calibration data.  While the MATLAB implementation is naturally expected to have a major disadvantage in this comparison, it is notable that the PISCO-based MATLAB computation completed in $\sim$1 second, which was substantially faster than the $\sim$19 seconds used by the C-based implementation.  This underscores the promise of PISCO, and strongly suggests that a C-based implementation of PISCO might be worthwhile.
 
 The authors have made an open-source implementation of the nullspace-based algorithm with PISCO available at http://mr.usc.edu/download/pisco/.

\bibliographystyle{IEEEtran}
\bibliography{./bibliography}

\end{document}